\documentclass[12pt]{article}
\pdfoutput=1
\usepackage{jheppub}
\usepackage{datetime}
\usepackage{float}
\usepackage{slashed}
\usepackage{multirow}
\usepackage{enumerate}
\usepackage{amsmath}
\usepackage{comment}
\usepackage{color}
\usepackage{amssymb}
\usepackage{pifont}
\usepackage[stable]{footmisc}
\usepackage{hyperref}

\newcommand{\be}{\begin{equation}}
\newcommand{\ee}{\end{equation}}
\newcommand{\bea}{\begin{eqnarray}}
\newcommand{\eea}{\end{eqnarray}}
\newcommand{\bean}{\begin{eqnarray*}}
\newcommand{\eean}{\end{eqnarray*}}

\newcommand{\bF}{\mathbb{F}}
\newcommand{\bX}{\mathbb{X}}
\newcommand{\bbX}{\overline{\mathbb{X}}}

\newcommand{\calI}{\mathcal{I}}

\title{Non-Renormalization For Non-Supersymmetric Black Holes}

\author{Anthony M. Charles,}
\author{Finn Larsen, and}
\author{Daniel R. Mayerson}

\affiliation{Department of Physics and Michigan Center for Theoretical Physics, \\
University of Michigan, 450 Church Street, Ann Arbor, MI 48109-1020, USA}

\emailAdd{amchar@umich.edu}
\emailAdd{larsenf@umich.edu}
\emailAdd{drmayer@umich.edu}

\abstract{
We analyze large logarithmic corrections to 4D black hole entropy and relate them to the Weyl anomaly. We use duality to show that counter-terms in 
Einstein-Maxwell theory can be expressed in terms of geometry alone, with no dependence on matter 
terms. We analyze the two known ${\cal N}=2$ supersymmetric invariants for various non-supersymmetric black holes and find that both reduce to 
the Euler invariant. The $c$-anomaly therefore vanishes in these theories and the coefficient of the large logarithms becomes topological. It is therefore independent of continuous 
black hole parameters,  such as the mass, even far from extremality. 
}

\begin{document}
\maketitle
\flushbottom

\section{Introduction and Summary}
Precision results for the entropy of BPS  black holes give detailed insights into the quantum structure of black holes (see e.g.~\cite{Strominger:1996sh,Maldacena:1997de,Ooguri:2004zv,Kraus:2005vz}). The techniques 
underlying these results involve extrapolation from weak to strong coupling of quantities that are known to be protected by supersymmetry. 
The physics of black holes with no supersymmetry is much more complicated and it is generally expected that precision results for their entropy 
is not possible. In this paper we present evidence that may indicate some precision studies of non-supersymmetric black holes are possible, after all: 
certain black holes satisfy a non-renormalization theorem when they are embedded in theories with ${\cal N} = 2$ supersymmetry even though the 
black holes themselves do not preserve any supersymmetry, not even an approximate supersymmetry. Moreover, our non-renormalization theorem 
is protected by a topological invariant. 

The objects we study are logarithmic quantum corrections to black hole entropy.  The leading order quantum corrections to the Bekenstein-Hawking area law scale with the logarithm of the black hole horizon area.  It is known that these large logarithms offer an infrared window into 
ultraviolet physics: they are computable in the low energy theory and yield precision data that must be matched by sub-leading terms in the asymptotic 
density of black hole microstates \cite{Sen:2011ba,Bhattacharyya:2012wz,Sen:2012dw}. Agreement with the microscopic theory has been established 
in those (highly supersymmetric) cases where precision counting is available \cite{Banerjee:2010qc,Banerjee:2011jp,Pathak:2016vfc}. 
We discuss these logarithms for non-supersymmetric black holes using effective quantum field theory. 
\\

The current work is developed with a particular setting in mind, previously discussed by two of us in \cite{Charles:2015eha}. We embed the standard Einstein-Maxwell gauge field $F_{\mu\nu}$ into $\mathcal{N}=2$ supergravity (with any number of vector multiplets, enumerated by the index $I$) as:
\be \label{eq:introEM} F^{+I}_{\mu\nu} = X^I F_{\mu\nu}^+~,\ee
where $X^I$ and $F_{\mu\nu}^I$ are (respectively) the scalar and vector field strength of a $\mathcal{N}=2$ vector multiplet, and the scalars $X^I$ are taken to be constant (see section \ref{sec:form} for more details). In this way, we can obtain non-supersymmetric solutions in $\mathcal{N}=2$ supergravity, such as non-extremal Kerr-Newman black holes. Fluctuations of the ${\cal N}=2$ matter exhibit non-minimal couplings in this environment which, by explicit computation, were found to modify the Weyl anomaly coefficients from their standard values such that the total central charge $c=0$ for a complete ${\cal N}=2$ multiplet. The present paper complements the explicit computations in \cite{Charles:2015eha} by 
explaining how the null result follows from symmetries, effectively proving a non-renormalization theorem for these non-supersymmetric solutions in $\mathcal{N}=2$ supergravity.

We prove our non-renormalization theorem by exploiting several symmetries which heavily constrain the effective quantum field theory of quantum corrections to black holes.  The analysis of each of these symmetries encounters conceptual questions that we address:
\begin{itemize}
\item
{\bf Scaling symmetry}: classical gravity is not a conformal theory, yet it is conventional to express one-loop quantum corrections in terms of a Weyl anomaly in the trace of the energy momentum tensor\footnote{The notation $E_4$ and $W^2$ is given explicitly in section~\ref{sec:effQFT}.}~\cite{Christensen:1978md,Christensen:1979iy,Birrell:1982ix,Vassilevich:2003xt}:
\begin{equation}
T_\mu^{~\mu}   = \frac{1}{16\pi^2} \left( aE_4 - cW^2 + \ldots \right)~.
\label{eqn:traceanomaly1}
\end{equation}
We discuss this terminology in the language of effective quantum field theory and relate it to the logarithmic corrections to black hole entropy. 
\item
{\bf Duality}: the equations of motion of classical electrodynamics are invariant under electromagnetic duality but the corresponding classical action is not \cite{Gaillard:1981rj}. 
We show that duality constrains the dependence of the quantum 
action on the explicit field strength and, in the case of Einstein-Maxwell theory, eliminates it entirely.  In this case the dots indicating additional terms in the trace anomaly 
(\ref{eqn:traceanomaly1}) are absent and the effect of matter has been entirely absorbed into the values of the coefficients $a, c$ which then take non-standard values. 
\item
{\bf Supersymmetry}: for black hole solutions to theories with ${\cal N}=2$ supersymmetry the quantum effective action is constrained by on-shell supersymmetry. In $D=4$ there 
are two known distinct four derivative invariants \cite{Butter:2013lta,Butter:2014iwa}. They complete the two terms written explicitly in (\ref{eqn:traceanomaly1}) with particular matter terms and take the schematic form
\begin{equation}
E_4 + {\rm SUSY~matter}~,\quad W^2+ {\rm SUSY~matter}~,
\label{eqn:fourderivsusy}
\end{equation}
in an off-shell formalism. We show that, when evaluated on-shell for our class of solutions (\ref{eq:introEM}), both ${\cal N}=2$ invariants reduce to just the Euler invariant $E_4$. 
Thus supersymmetry excludes the second term in the trace anomaly (\ref{eqn:traceanomaly1}), so $c=0$. 

The significance of this result is that the logarithmic correction to black hole entropy reduces to a topological quantity, independent of the black hole parameters. In particular, it can be deformed from the extremal (supersymmetric) limit to a generic (non-supersymmetric) black hole without any change in value. This property suggests an underlying index 
theorem, a great surprise in the context of non-supersymmetric black holes. 
\end{itemize}

Our results may superficially appear in conflict with findings obtained in some other areas of inquiry. 
For example, physical principles require the ratio $c/a\sim 1$ for conformal field theory in a curved background, with precise ``conformal collider" bounds easily 
excluding $c=0$ \cite{Hofman:2008ar,Hofman:2016awc,Afkhami-Jeddi:2016ntf}. Such apparent conflicts are simply due to the additional matter contributions that arise when we take dynamical gravity into account. Our considerations are thus consistent with standard results and complementary to several areas of recent research. 

The most obvious generalization of our work would be to understand whether the class of non-supersymmetric solutions (\ref{eq:introEM}) for which our non-renormalization theorem $c=0$ holds can be broadened and generalized further. In particular, it would be interesting to analyze solutions with non-constant scalars. However, as we discuss, the possible four-derivative corrections to more general backgrounds are expected to involve more (and more complicated) supersymmetric invariants, especially when the scalars are not constant.  This will require the introduction of new four-derivative supersymmetric invariants beyond the two we consider. 

Our calculations derive a $c=0$ non-renormalization theorem from the symmetries of $\mathcal{N}=2$ supergravity. It would be interesting to understand the $c=0$ result from the different perspective of a (super-)index theorem in the spirit of other gravitational indices (such as e.g. \cite{Christensen:1978md}). The strategy employed to establish such theorems involve relating quadratic fluctuations of bosons around the background to those of fermions.  When the non-zero modes can be shown to cancel, the only contribution to the quantum corrections comes from the zero modes and is thus topological. There are many examples where this mechanism applies but they generally rely on supersymmetry preserved by the background.  It would be novel if index theorems can be generalized to non-supersymmetric backgrounds such as ours. If it is possible it might also help understand how and when one could generalize our non-renormalization theorem to a broader class of solutions.

As stressed in the opening, an important motivation for this work is the potential for a microscopic understanding of black hole entropy and quantum corrections to it. 
Detailed microscopic models have been established for various types of supersymmetric black holes, using tools inherent to supersymmetry. Analogous microscopic 
descriptions of non-supersymmetric black holes are typically elusive and, if known, difficult to handle. Our work identifies a family of non-supersymmetric black holes that 
enjoys a simple and restricted form of one-loop quantum corrections because they are solutions in a theory with supersymmetry. This suggests an underlying 
structure that may point toward a microscopic description of such non-supersymmetric black holes.
\\

The rest of this paper is organized following the three bullet points above: sections~\ref{sec:effQFT}, \ref{sec:duality}, and \ref{sec:resultssusy} address scaling symmetry, duality, and supersymmetry in turn, with an interlude in section~\ref{sec:form} to briefly summarize relevant details of ${\mathcal N}=2$ supergravity formalism and the particular class of solutions considered. Several appendices review further details, especially of off-shell formalism for ${\mathcal N}=2$ supergravity.

\section{Effective Quantum Field Theory}\label{sec:effQFT}
In this section we formulate the computation of logarithmic corrections to black hole entropy in the framework of effective quantum field theory. The purpose is to 
connect results from Euclidean quantum gravity developed by Sen \cite{Sen:2012dw} with other approaches. Since the material in this section 
is not really new we focus on conceptual issues and, for clarity, we limit ourselves to four spacetime dimensions.

\subsection{Scaling Transformations for Gravity}
A key ingredient of effective quantum field theory is simple dimensional analysis, exploited to order scales in the problem according to their importance. To have a representative example in mind consider Einstein gravity minimally coupled to a vector and a scalar field:
\begin{equation}
I [ g_{\mu\nu}, A_\mu, \phi ]  =  \int d^4 x\sqrt{-g}\,{\cal L}  = \frac{1}{ 2\kappa^2} \int d^4 x \sqrt{-g}\left( R - \frac{1}{4} F_{\mu\nu} F^{\mu\nu} - \frac{1}{2} \partial_\mu \phi \partial^\mu \phi \right) ~.
\label{eqn:twoderaction}
\end{equation}
This action transforms under a rigid scaling as a homogeneous function of degree two: 
\begin{equation}
I [ \lambda^2 g_{\mu\nu}, \lambda A_\mu, \phi ]  = \lambda^2 I [ g_{\mu\nu}, A_\mu, \phi ] ~.
\label{eqn:scaling}
\end{equation}
The scaling dimensions in this formula are different from the usual mass dimensions in quantum field theory where, for example, $A_\mu$ and $\phi$ both have 
mass dimension one. It is also different from Weyl rescaling which, for example, would be a true symmetry of the pure Maxwell term and not just homogeneity. 
On the other hand, the homogeneity of the gravitational action is very well-known in the context of black holes. 
For example, 
the Smarr relation $2S = \beta M - \beta \Phi Q -  2\beta \Omega J$ for thermodynamic variables is equivalent to an entropy with homogeneity of degree 
two 
\begin{equation}
S [ \lambda M , \lambda^2 J, \lambda Q ]  = \lambda^2 S [ M , J, Q ]~,
\label{eqn:entropyscaling}
\end{equation}
which is equivalent to the corresponding property (\ref{eqn:scaling}) of the action. In this paper we consider only theories with matter and couplings that respect homogeneity of degree two at the classical level. 

In addition to terms with the structure introduced in (\ref{eqn:twoderaction}), homogeneity of degree two is also respected by the Pauli couplings that are characteristic of fermions in supergravity such as the ${\cal N}=2$ gravitini
\begin{equation}
I_{\rm gravitini}  
= - \frac{1}{2\kappa^2} \int d^4 x \sqrt{-g} \left[ 2{\bar\psi}_{i\mu}\gamma^{\mu\nu\rho} \nabla_\nu \psi^i_\rho + (F_{\mu\nu}^- {\bar\psi}^\mu_i\psi^\nu_j\epsilon^{ij} + {\rm h.c.}) \right]~,
\label{eqn:gravitini}
\end{equation}
if we assign scaling factor $\lambda^{1/2}$ to $\psi^i_\mu$. 
On the other hand, the scaling symmetry (\ref{eqn:scaling}) is violated by a minimal coupling to a gauge field and also by a potential for the scalar field(s).

\subsection{Background Field Formalism}
It will come as no surprise that the scaling relation enjoyed by the classical theory is violated by quantum effects since scaling violation is known from 
any textbook in quantum field theory. However, the gravitational setting is less well-known, so it is worth discussing somewhat pedagogically
in the familiar language of effective quantum field theory. 

We employ the background field method and introduce the quantum effective action
\begin{equation}
\Gamma [ g_{\mu\nu}, A_\mu, \phi ] = - i \log \int [\mathcal{D}\delta g_{\mu\nu} \mathcal{D}\delta A_\mu \mathcal{D}\delta \phi] e^{iI [ g_{\mu\nu} + \delta g_{\mu\nu}, A_\mu + \delta A_\mu, \phi +\delta\phi]}~.
\label{eqn:quantumaction}
\end{equation}
Although we have not indicated them explicitly, the action $I$ must include gauge-fixing terms that impose background field gauge
\begin{equation}
\nabla^\mu \delta A_\mu = 0~,\quad \nabla^\mu \delta g_{\mu\nu} = 0 ~,
\label{eqn:gaugecondition}
\end{equation}
on the fluctuating fields. Gauge-fixing preserves homogeneity of degree two but it breaks diffeomorphism invariance and gauge symmetry of the fluctuating fields 
so that the path integral can be performed. Importantly, in 
background field gauge (\ref{eqn:gaugecondition}) the quantum effective action $\Gamma$ is nonetheless invariant under both diffeomorphisms and $U(1)$ gauge 
symmetry acting on background fields: 
it realizes symmetries explicitly. 

The usual quantum effective action is the generator of the 1PI diagrams and so its tree-level amplitudes give the complete 
amplitudes of the full quantum theory. As such it is closely related to the physical S-matrix that encodes all scattering amplitudes in flat 
space. The quantum effective action  (\ref{eqn:quantumaction}) we analyze is formally the same as this standard object, but we expand around a black hole background rather than flat space. The black 
hole is not merely a deformation of flat space by a source: it has nontrivial Euler number 
\begin{equation}
\chi = \frac{1}{32\pi^2} \int d^4 x \sqrt{-g} E_4  = 2 ~,
\label{eqn:eulerinvariant}
\end{equation}
where the Gauss-Bonnet invariant is
\begin{equation}
E_4 = R_{\mu\nu\rho\sigma}R^{\mu\nu\rho\sigma} - 4R_{\mu\nu} R^{\mu\nu}  + R^2  ~.
\label{eqn:gb}
\end{equation}
Thus a black hole and flat space are in distinct distinct topological sectors. The black hole is similar to a soliton or an instanton. 
The quantum effective action in this setting is conceptually different from the S-matrix in flat space. Indeed, the non-trivial Euler number (\ref{eqn:eulerinvariant}) makes 
it more interesting. For example, there is a one loop contribution even in pure gravity (some recent discussions include \cite{BjerrumBohr:2002kt,Bern:2015xsa}).  

\subsection{Regularization and Renormalization}
As usual, explicit computation of the effective action identifies divergences that must be regularized and removed through renormalization. 
Multiple scales appear in this process: 
\begin{itemize}
\item
{\it The physical scale $M$}: we assume for simplicity that all relevant scales of the black hole are 
comparable $M\sim M_{\rm BH}$. Then the curvature $R\sim (GM)^{-2}$ and the field strength 
$F\sim (GM)^{-1}$. Generic black holes and extremal black holes each have just one scale, up to ratios of ${\cal O}(1)$, so they satisfy our assumption. However, the interpolation between these cases necessarily introduces a large dimensionless ratio so it involves a parametrically larger length scale (such as the thermal wave length). 
We do not study such transitions. 
\item
{\it The cutoff scale $\Lambda\gg M$}: the upper limit to the validity of the low energy effective field theory in the Wilsonian picture. This is the reference scale where couplings in effective field theory are defined through boundary conditions generated by matching with the UV theory. 
\item
{\it The UV scale $\Lambda_{\rm UV}\gg\Lambda$}: 
typical masses of string states or other high energy excitations. In effective quantum field theory UV dynamics has been integrated out so 
that the UV scale does not appear in the low energy action. Indeed, physics at the UV scale can be decoupled by taking the limit $\Lambda_{\rm UV}/\Lambda\to\infty$ with couplings at the cutoff scale $\Lambda$ kept fixed. 
\end{itemize} 
Additionally, since black holes in asymptotically flat space are unstable there is also need for an IR cutoff cutoff $\Lambda_{\rm IR}\ll M$ that regulates the nominally infinite volume of spacetime. It will play no role in our discussion.

\subsection{Local Terms in the Action}
The quantum effective action for general gravitational backgrounds is very complicated and 
non-local (for some recent discussions, see \cite{Godazgar:2016swl,Bautista:2015nxc}). We are just interested in its transformation under rigid scale 
transformations and so it is sufficient to consider terms that are local except for their possible scale dependence. 
The leading terms in the effective action are the two-derivative terms that appear already in the (schematic) classical action (\ref{eqn:twoderaction}), but there are also corrections incorporated in higher derivative operators such as
\begin{equation}
{\cal O}_{2n} = R^{n}~,~~R^{n-1} F^{2}~,~~R^{n-1} (\partial\phi)^2~,~ \ldots~,
\label{eqn:higherdim}
\end{equation}
with $n\geq 2$ and various contractions of indices implied. Dimensional analysis determines the typical contribution from any of the ${\cal O}_{2n}$ 
operators as
\begin{equation}
c_{2n} {M^{2n-4}\over \Lambda^{2n-4}}~,
\label{c2ndef}
\end{equation}
where $c_{2n}$ is a numerical constant. Generally they are parametrically small at energies far below the cutoff $M\ll\Lambda$, {\it i.e.} 
for large black holes. However, there is an exception for the marginal operators $n=2$, {\it i.e.} those with four derivatives 
\begin{equation}
{\cal O}_4 = R^2~,~~R F^2~,~~R (\partial\phi)^2 ~,~ \ldots
\label{eqn:fourder}
\end{equation}
For $n=2$ dimensional analysis is not only consistent with a finite effect 
\begin{equation}
c_{2n} \frac{M^{2n-4}}{\Lambda^{2n-4}}\sim c_4~,
\label{c4def}
\end{equation}
but it also allows a logarithm 
\begin{equation}
c_4'\log {M\over\Lambda}~. 
\label{eqn:largelog}
\end{equation}
This logarithm is potentially large, since we assume $M\gg\Lambda$, and it is entirely 
due to the light fields retained below the cutoff, typically the massless fields, so its coefficient is computable without knowledge of the UV theory. 
Moreover, in the gravitational setting the coefficient of the large logarithm is determined exactly by a one-loop computation: the coefficient of ${\cal O}_{2n}$ operators receive
contributions only at the $(n-1)$th loop order because the gravitational coupling constant $\kappa\sim M^{-1}_{\rm pl}$ has mass dimension $[\kappa]=-1$. 

It is disconcerting that the large logarithm (\ref{eqn:largelog}) depends on the arbitrary cut-off $\Lambda$. This should be understood together with the fact that 
the marginal operators (\ref{eqn:fourder}) all appear with a finite dimensionless coefficient already in the effective field theory defined at the cut-off, 
before quantum corrections are included. 
This ``primordial" coefficient, denoted by $c_{4}$ in (\ref{c4def}), becomes a running coupling in the quantum theory $c_4 \to c_{4}(M) = c_4 + c_4'\log {M\over\Lambda}$.
Alternatively, dimensional transmutation allow us to introduce the dynamical scale  
\begin{equation}
\Lambda_{\rm dyn} = M e^{-{c_{4}(M)\over c_4'}}~,
\label{eqn:dynamical}
\end{equation}
and absorb the finite piece in the large logarithm
\begin{equation}
c_{4}(M) = c_4'\log {M\over{\Lambda_{\rm dyn}}}~.
\end{equation}
This represents a conceptual advantage because the dynamical scale (\ref{eqn:dynamical}) is RG invariant, {\it i.e.} independent of the arbitrary cut-off scale $\Lambda$. The dynamical scale (\ref{eqn:dynamical}) is a non-perturbative scale intrinsic to quantum gravity \cite{Tong:2014era}. 
More precisely, there can in general be multiple dynamical scales, one for each term in (\ref{eqn:fourder}).

The ``primordial" coefficient $c_{4}$ is ambiguous because it depends on the renormalization scheme. This ambiguity is not a concern because the constant term is 
anyway dominated by the large logarithm which is scheme independent. Since this quantum contribution is generated by the light modes it will respect symmetries of 
the low energy theory, such as duality symmetry, whether or not those symmetries are preserved by the UV theory. We will exploit this feature in sections~\ref{sec:duality} and~\ref{sec:resultssusy}. 

\subsection{The Weyl Anomaly}
A common quantitative measure of the logarithmic corrections is the contribution of these classically marginal operators to the
trace of the renormalized energy momentum tensor
\begin{equation}
\int d^4 x \sqrt{-g}\left( T_\mu^{~\mu} \right)_{\rm an} \equiv  \left( M\partial_M - 2\right) \Gamma = {1\over 16\pi^2} \int d^4 x \sqrt{-g}\left(  aE_4- cW^2 + \ldots \right)~,
\label{eqn:trace}
\end{equation}
where $E_4$ is the Gauss-Bonnet invariant (\ref{eqn:gb}), the square of the Weyl tensor is
\begin{equation}
W^2 = R_{\mu\nu\rho\sigma}R^{\mu\nu\rho\sigma} - 2R_{\mu\nu} R^{\mu\nu}  + {1\over 3} R^2  ~,
\end{equation}
and the dots denote other four derivative operators, including those formed from matter fields. The notation is adopted from the scale anomaly of conformal field theory in non-dynamical 
background geometries but the physical interpretation is different here. The classical action is not invariant under scaling, 
it transforms homogeneously with dimension two (\ref{eqn:scaling}). It is this homogeneity that is violated by the quantum effective action. 
The scaling parameter $\lambda$ introduced in (\ref{eqn:entropyscaling}) is identified as $\lambda = \Lambda/M$. 

The values of the numerical coefficients $a$ and $c$ appear in many physical applications. The simplest fields couple just to gravity and then the matter 
terms denoted by dots in (\ref{eqn:trace}) are absent. In this case the anomaly coefficients are well-known~\cite{Christensen:1978md,Birrell:1982ix,Christensen:1978gi}. The simplest are collected in table \ref{table:simpleac}. 
\bgroup
\def\arraystretch{1.5}
\begin{table}[H]
\centering
\begin{tabular}{|c|c|c|}
\hline
\textbf{Field} & $c$ & $a$ \\ \hline
Complex Scalar & $\frac{1}{60}$ & $\frac{1}{180}$ \\ \hline
Weyl Fermion & $\frac{1}{40}$ & $\frac{11}{720} $ \\ \hline
Vector  & $\frac{1}{10}$ & $\frac{31}{180}$ \\ \hline
Gravitino & $-\frac{411}{360}$  &  $-\frac{589}{720}$ \\ \hline
\end{tabular}
\caption{Central charges $c$ and $a$ for neutral fields with minimal coupling to gravity. Each entry has two physical degrees of freedom.}
\label{table:simpleac}
\end{table}
\egroup

However, in order for the effective action to allow a black hole as a consistent semiclassical saddle-point, we must treat gravity as dynamical. The geometry cannot be 
selected freely, it must be related by Einstein's equation to matter that, in turn, is also on-shell. The only situation where matter can be omitted altogether is when the 
background is Ricci-flat $R_{\mu\nu}=0$, such as for the Kerr black hole\footnote{In this paper we do not consider the possibility that the ``matter" is a cosmological constant. The $c$ and $a$ coefficients for this case was discussed in \cite{Larsen:2015aia}.}. That case is so special that the $c$ and $a$ coefficients cannot even be distinguished,
since $E_4$ and $W^2$ both reduce to $R_{\mu\nu\rho\sigma}R^{\mu\nu\rho\sigma}$. In the more general case where we do allow for appropriate 
matter there will necessarily be additional terms in the effective action: the dots in (\ref{eqn:trace}) become non-trivial. 

Generally, there can be numerous matter terms and the effective action may be very involved. Our focus is on some special cases where the theory is essentially
Einstein-Maxwell (gravity coupled to electromagnetism). We consider this theory by itself and also as a subsector of ${\cal N}\geq 2$ supergravity coupled to additional vector and 
hyper multiplets. In all those cases, the matter terms in the effective action combine into a special form such that, when Einstein's equation is imposed, they can be absorbed into the purely geometrical terms, {\it i.e.} they shift  $c$ and $a$ from their standard values. In section~\ref{sec:duality} we use duality symmetry to show why this is possible.

\bgroup
\def\arraystretch{1.5}
\begin{table}[H]
\centering
\begin{tabular}{|c|c|c|}
\hline
\textbf{Fields} & $c$ & $a$ \\ \hline
Fermions in ${\cal N}=2$ Hypers & $ - \frac{1}{30}$ & $-\frac{19}{360}$  \\ \hline
Vectors/Scalars in ${\cal N}=2$ Vectors  & $-\frac{1}{20}$ & $\frac{1}{90}$ \\ \hline
Gravitino in ${\cal N}>2$ Supergravity & $-\frac{1}{5}$  &  $\frac{41}{360}$ \\ \hline
Graviton/R-vector in ${\cal N}=2$ Supergravity & $\frac{411}{180}$  &  $\frac{106}{90}$ \\ \hline
\end{tabular}
\caption{Central charges $c$ and $a$ for some nontrivial fields in $\mathcal{N} \geq 2$ supergravity theory. Each entry has a total of four degrees of freedom.}
\label{table:N=2}
\end{table}
\egroup
In previous work two of us explicitly computed the values for $c$ and $a$ for various multiplets in ${\cal N}=2$ supergravity~\cite{Charles:2015eha}. In each type of ${\cal N}=2$ multiplet 
(hyper, vector, gravitino, gravity) either the bosons or the fermions are minimally coupled but the fields with opposite statistics are subject to non-minimal matter couplings
that lead to shift of the $c$ and $a$ coefficients. The results for the fields with non-minimal couplings are given in table \ref{table:N=2}. It is striking that when these components are completed 
into full ${\cal N}=2$ multiplets by adding 
appropriate minimally coupled fields from table \ref{table:simpleac}, the total value of the $c$ central charge vanishes in each of four cases. 
In section~\ref{sec:resultssusy} we show that off-shell supersymmetry is responsible for this result.

 \subsection{The Black Hole Entropy}
\label{sec:effQFT:entropy}
 
The quantum effective action is not a conventional observable but it is closely related to the black hole entropy. With the provisional identification
$\delta S = - \delta \Gamma$ for the quantum correction to the black hole entropy the Weyl anomaly (\ref{eqn:trace}) gives
\begin{equation}
\delta S_{\rm continuum} =  - {1\over 16\pi^2} \int d^4 x\left(  aE_4- cW^2 + \ldots \right)\log {M\over{\Lambda_{\rm dyn}}} ~.
\label{eqn:entropy}
\end{equation}
In gravitational physics it is conventional to relate the physical scale to the horizon area $A_H \sim (GM)^2$ but in effective quantum field theory it is more natural to retain $M$. 

In the simplest case where either $c=0$ or the background is conformally flat (such as in AdS$_2\times S^2$) the Euler number (\ref{eqn:eulerinvariant}) 
for a black hole gives simply
\begin{equation}
\delta S_{\rm continuum} =  - 4 a \log {M\over{\Lambda_{\rm dyn}}} ~.
\end{equation}
In more general cases the $W^2$ term introduces a complicated dependence on black hole parameters~\cite{Henry:1999rm,Cherubini:2003nj}.

The identification $\delta S = - \delta \Gamma$ is appropriate in the microcanonical ensemble but in Euclidean quantum gravity it is more natural to impose
thermal boundary conditions. Then the on-shell action becomes identified with the free energy (multiplied by the inverse temperature) and to obtain the entropy one must Legendre 
transform. Such changes of ensemble modifies the coefficient of the large logarithm in the entropy (\ref{eqn:entropy}). 
Specifications of ensemble for the angular momentum and for electric/magnetic charges may similarly shift the coefficient. Moreover, standard values for the $a$ and $c$ 
anomaly coefficients are computed for generic geometry while, for black hole backgrounds, there are zero-modes due to global symmetries and those also contribute to the 
large logarithm. The total contribution from all these discrete effects takes the form
\begin{equation}
\delta S_{\rm discrete} = n \log {M\over{\Lambda_{\rm dyn}}}~,
\end{equation}
where $n$ is an integer. The value of $n$ depends on ensemble and global symmetries but it is independent of the matter content of the theory and also does not depend on 
continuous black hole parameters (except for jumps at extremal limits). For example, $n=2$ for a BPS black hole and $n=-1$ for a generic 
Kerr-Newman black hole (with $J_3$ fixed and $\vec{J}^2$ arbitrary). Several other options were discussed by Sen \cite{Sen:2012dw} and there is a summary of various 
situations in \cite{Charles:2015eha}. 

The Euclidean quantum gravity path integral with thermal boundary conditions is also sensitive to the contribution from a thermal gas in equilibrium with Hawking 
radiation from the black hole. The temperature of the black hole is $T\sim M$ so the on-shell action is $I \sim V T^3 \sim M^3/\Lambda^3_\text{IR}$ if the entire system is regulated by 
a large box with typical length scale $\sim \Lambda^{-1}_\text{IR}$. Lower powers of $M/\Lambda_\text{IR}$ could appear, due to boundary conditions at the box, but no dependence on the cut-off 
scale $\Lambda$ can appear in the thermal bath since the physical parameter $M$ and the IR cut-off $\Lambda_\text{IR}$ are both kept fixed as $\Lambda$ varies. Therefore, the 
contribution from the thermal gas can be removed from the finite part of the on-shell action without affecting the large logarithm depending on the ratio $M/\Lambda$.  
The same conclusion follows if we keep $\Lambda$ fixed and rescale the physical parameter $M$ using the classical transformation (\ref{eqn:scaling}) since then 
the IR cut-off $\Lambda_\text{IR}$ must be scaled as well, keeping the ratio $M/\Lambda_\text{IR}$ fixed. Either way, the IR cut-off $\Lambda_\text{IR}$ decouples from our considerations.


\section{Supergravity Formalism and Black Hole Solutions}
\label{sec:form}


Our results on duality in section~\ref{sec:duality} and supersymmetry in section~\ref{sec:resultssusy} make extensive use of both the off-shell and on-shell 
formulations of $\mathcal{N}=2$ supergravity. In this section, we review the essential parts needed to understand our methods and the relevant class of black hole solutions. More details of off-shell $\mathcal{N}=2$ supergravity are reviewed in Appendix \ref{sec:appN2}. 


\subsection{Field Content}
\label{sec:form:field}


The off-shell formalism realizes $\mathcal{N}=2$ supergravity in 4D by imposing constraints on superconformal multiplets whose fields transform under the $\mathcal{N}=2$ superconformal group.  The most important of these multiplets is the Weyl multiplet, which contains the gauge fields associated with each of the superconformal symmetry generators.  The independent fields in this Weyl multiplet are
\begin{equation}
	\left(\,e_\mu^{~a}~,~\psi_\mu^i~,~b_\mu~,~A_\mu~,~\mathcal{V}_{\mu~j}^{~i}~,~T^-_{\mu\nu}~,~\chi^i~,~D~\right)~,
	\label{eqn:Weylmultiplet}
\end{equation}
where $e_\mu^{~a}$ is the metric vierbein, $\psi^i_\mu$ is the gravitino, $b_\mu$ is the dilatation generator, $A_\mu$ is an auxiliary $U(1)_R$ gauge field, $\mathcal{V}_{\mu~j}^{~i}$ is an auxiliary $SU(2)_R$ gauge field, $T^-_{\mu\nu}$ is an auxiliary anti-self-dual tensor, $\chi^i$ is an auxiliary $SU(2)$ doublet of Majorana spinors, and $D$ is an auxiliary real scalar field.  The Weyl multiplet has $24+24$ bosonic and fermionic degrees of freedom off-shell.

We will introduce matter in the form of $n_V +1$ off-shell $\mathcal{N}=2$ vector multiplets, denoted by $\mathbf{X}^I$ where $I = 0,\ldots, n_V$.  These will reduce down to $n_V$ physical vector multiplets in the on-shell theory.  The field content of the vector multiplets is
\begin{equation}
	\mathbf{X}^I = \left(\,X^I~,~\Omega^I_i~,~W_\mu^I~,~Y^I_{ij}~\right)~,
\label{eq:vectormult}
\end{equation}
where $X^I$ is a complex scalar, $\Omega^I_i$ is an $SU(2)$ doublet of chiral gauginos, $W_\mu^I$ is a $U(1)$ vector gauge field, and $Y^I_{ij}$ is an auxiliary $SU(2)$ triplet of real scalars.  Each vector multiplet has $8+8$ degrees of freedom off-shell.  The scalars $X^I$ have Weyl weight $w=1$ and $U(1)_R$ charge (referred to as a chiral weight) 
$c = -1$, while their Hermitian conjugates $\bar{X}^I$ have the same Weyl weight and opposite chiral weight.  The vector fields $W_\mu^I$ are uncharged under the $U(1)_R$ symmetry.

The field strengths of the auxiliary $U(1)_R$ gauge field $A_\mu$ and the auxiliary $SU(2)_R$ gauge field $\mathcal{V}_{\mu~j}^{~i}$ are (respectively)
\begin{equation}
	A_{\mu\nu} \equiv \partial_\mu A_\nu - \partial_\nu A_\mu~,
\end{equation}
\begin{equation}
	\mathcal{V}_{\mu\nu~j}^{~~i} \equiv \partial_\mu \mathcal{V}_{\nu~j}^{~i} - \partial_\nu \mathcal{V}_{\mu~j}^{~i} + \frac{1}{2}\mathcal{V}_{\mu~k}^{~i}\mathcal{V}_{\nu~j}^{~k} - \frac{1}{2}\mathcal{V}_{\nu~k}^{~i}\mathcal{V}_{\mu~j}^{~k}~.
\end{equation}
The field strengths of the vector multiplet gauge fields are
\begin{equation}
	F^I_{\mu\nu} \equiv \partial_\mu W^I_\nu - \partial_\nu W_\mu^I~.
\end{equation}
We will also make use of the supercovariant field strengths
\begin{equation}\label{eq:curlyFdef}
	\mathcal{F}^{-I}_{\mu\nu} \equiv F^{-I}_{\mu\nu} - \frac{1}{4}\bar{X}^I T^-_{\mu\nu}~, \quad \mathcal{F}^{+I}_{\mu\nu} \equiv F^{+I}_{\mu\nu} - \frac{1}{4}X^I T^+_{\mu\nu}~,
\end{equation}
where $F^{\pm I}_{\mu\nu}$ are the (anti-)self-dual parts of the vector multiplet field strengths as defined in (\ref{eq:N2:Fpm}).


\subsection{Two-Derivative Theory}
\label{sec:form:twoderiv}


The couplings between the vector multiplets and the Weyl multiplet can be specified succinctly by a prepotential
\begin{equation}
	F = F^{(0)}(X^I)~,
\label{eq:prepot_2}
\end{equation}
a meromorphic function of the complex scalars in the vector multiplets. Its derivatives are denoted:
\begin{equation}
	F_I \equiv \frac{\partial F}{\partial X^I}~, \quad F_{\bar{I}} \equiv \frac{\partial F}{\partial \bar{X}^I} = 0~,
\end{equation}   
where the vanishing of the anti-holomorphic derivative follows from holomorphy. The prepotential is homogeneous with degree two under Weyl transformations. The vector multiplet scalars have Weyl weight one so $F^{(0)}$ must satisfy
\begin{equation}
	F^{(0)}(\lambda X^I) = \lambda^2 F^{(0)}(X^I)~.
\end{equation}
The two-derivative Lagrangian that couples the vector and Weyl multiplets via the prepotential (\ref{eq:prepot_2}) is
\begin{equation}\begin{aligned}
	8\pi\mathcal{L} &= \bigg{[} i \mathcal{D}^\mu F^{(0)}_I \mathcal{D}_\mu \bar{X}^I - i F^{(0)}_I \bar{X}^I\left(\frac{1}{6}R - D \right) -\frac{i}{8}F^{(0)}_{IJ}Y_{ij}^I Y^{Jij} \\
	&\quad + \frac{i}{4}F^{(0)}_{IJ}\mathcal{F}^{-I}_{\mu\nu}\mathcal{F}^{-\mu\nu J} - \frac{i}{8} F^{(0)}_I \mathcal{F}^{+I}_{\mu\nu}T^{+\mu\nu} - \frac{i}{32}F^{(0)} T_{\mu\nu}^+ T^{+\mu\nu}\bigg{]} +\text{h.c.} \\
	&\quad + \text{(fermions)}~.
\label{eq:lg}
\end{aligned}\end{equation}

We can reduce the superconformal symmetry to a Poincar\'e symmetry and further simplify the theory by imposing a consistent truncation
\begin{equation}\label{eq:trunc}
	b_\mu = Y_{ij}^I = \mathcal{V}_{\mu~j}^{~i} = \text{fermions} = 0~, \quad D = -\frac{1}{3}R~, \quad i(F^{(0)}_I \bar{X}^I - \bar{F}^{(0)}_I X^I) = \frac{8\pi}{\kappa^2}~.
\end{equation}
More details are reviewed in appendix~\ref{sec:appN2:trunc}.  Under this truncation, the two-derivative Lagrangian (\ref{eq:lg}) becomes
\begin{equation}\begin{aligned}
	\mathcal{L}^{(2)} &= -\frac{1}{2\kappa^2}R + \frac{1}{8\pi}\bigg{[} i \mathcal{D}^\mu F^{(0)}_I \mathcal{D}_\mu \bar{X}^I + \frac{i}{4}F^{(0)}_{IJ}\mathcal{F}^{-I}_{\mu\nu}\mathcal{F}^{-\mu\nu J} \\
	&\quad - \frac{i}{8} F^{(0)}_I \mathcal{F}^{+I}_{\mu\nu}T^{+\mu\nu} - \frac{i}{32}F^{(0)} T_{\mu\nu}^+ T^{+\mu\nu}\bigg{]} +\text{h.c.}~.
\label{eq:l_2deriv}
\end{aligned}\end{equation}
In the truncation (\ref{eq:trunc}), the supercovariant derivative acts on the scalar fields by
\begin{equation}
	\mathcal{D}_\mu X^I = \left(\partial_\mu + i A_\mu \right)X^I~.
\end{equation}
Thus the auxiliary fields $T^-_{\mu\nu}$ and $A_\mu$ both appear algebraically in the Lagrangian. Their equations of motion can be solved, yielding
\begin{equation}
\label{eq:Ttwoder}		T^-_{\mu\nu} = 4 \frac{N_{IJ} \bar{X}^J F^{-I}_{\mu\nu}}{N_{KL}\bar{X}^K \bar{X}^L}~, \quad A_\mu = i \frac{N_{IJ} \bar{X}^J \partial_\mu X^I}{N_{KL}\bar{X}^K X^L}~,
\end{equation}
where we have defined the Hermitian symplectic matrix $N_{IJ}$ as
\begin{equation}
	N_{IJ} = 2 {\rm Im} F^{(0)}_{IJ} ~.
\end{equation}
Eliminating the auxiliary fields $T^-_{\mu\nu}$ and $A_\mu$ from the action yields the bosonic terms in the familiar $\mathcal{N}=2$ supergravity Lagrangian
\begin{equation}\label{eq:lgonshell}
	\mathcal{L} = -\frac{1}{2\kappa^2}R - \frac{1}{8\pi}\mathcal{M}_{I \bar{J}} \partial^\mu X^I \partial_\mu \bar{X}^J - \frac{i}{32\pi}\mathcal{N}_{IJ}F^{+I}_{\mu\nu}F^{+\mu\nu J} + \text{h.c.}~,
\end{equation}
where the matrices $\mathcal{M}_{I\bar{J}}$ and $\mathcal{N}_{IJ}$ are defined by
\begin{equation}
		\mathcal{M}_{I\bar{J}} = N_{IJ} - \frac{N_{IK}\bar{X}^K N_{JL}X^L}{N_{MN}\bar{X}^M X^N}~, \quad \mathcal{N}_{IJ} = \bar{F}_{IJ}^{(0)} +i \frac{N_{IK}X^K N_{JL}X^L}{N_{MN}X^M X^N}~.
	\end{equation}
The Einstein, Maxwell, and Bianchi equations of the simplified theory are
\begin{align}
\label{eq:twoderEinstein}	R_{\mu\nu} &= -\frac{\kappa^2}{4\pi} \mathcal{M}_{I\bar{J}}\partial_{(\mu} X^I \partial_{\nu)} \bar{X}^J - \frac{i\kappa^2}{8\pi}\mathcal{N}_{IJ}F^{-I}_{\mu\rho}F^{+\rho J}_{\nu} + \text{h.c.}~, \\
\label{eq:twoderFEOM}	0 &= \nabla_\mu \left(\mathcal{N}_{IJ}F^{+\mu\nu J} - \bar{\mathcal{N}}_{IJ}F^{-\mu\nu J}\right)~, \\
\label{eq:twoderFBI}	0 &= \nabla_\mu \left( F^{+\mu\nu I} - F^{- \mu\nu I}\right)~.
\end{align}
These are the equations of motion for ${\cal N}=2$ supergravity.


\subsection{Four-Derivative Theory}
\label{sec:form:fourderiv}


Our main interest is to constrain the form of the Weyl anomaly (\ref{eqn:trace}), which contains four-derivative terms.  We therefore need to introduce higher-derivative corrections to the Lagrangian (\ref{eq:l_2deriv}).

Higher-derivative terms can be constructed in the off-shell $\mathcal{N}=2$ supergravity formalism by additionally coupling the theory to a chiral multiplet $\hat{\mathbf{A}}$.  The field content of the chiral multiplet is
\begin{equation}
	\hat{\mathbf{A}} = \left(\,\hat{A}~,~\hat{\Psi}_i~,~\hat{B}_{ij}~,~\hat{F}^-_{\mu\nu}~,~\hat{\Lambda}_i~,~\hat{C}~\right)~,
\label{eq:chiralmult}
\end{equation}
where $\hat{A}$ and $\hat{C}$ are complex scalars, $\hat{\Psi}_i$ and $\hat{\Lambda}_i$ are both $SU(2)$ doublets of left-handed fermions, $\hat{B}_{ij}$ is a complex $SU(2)$ triplet of scalars, and $\hat{F}^-_{\mu\nu}$ is an anti-self-dual tensor.  A chiral multiplet can have any Weyl weight $w$ from which the Weyl and chiral weights of the component fields can be determined.  In particular, the scalars $\hat{A}$ and $\hat{C}$ have Weyl weights $w$ and $w + 2$ and chiral weights $-w$ and $-w+2$, respectively.

The chiral multiplet will eventually be realized as a composite of the Weyl and vector multiplets such that four-derivative terms are introduced into the action.  The truncation (\ref{eq:trunc}) can be augmented by setting all fermionic and $SU(2)_R$-charged chiral multiplet fields to zero:
\begin{equation}
	\hat{\Psi}_i = \hat{\Lambda}_i = \hat{B}_{ij} = 0~.
\label{eq:form:chiraltrunc}
\end{equation}

The prepotential $F$ still determines all couplings in the theory but, in order to introduce higher-derivative interactions, it must be modified to become a function of the chiral multiplet scalar $\hat{A}$ as well as the vector multiplet scalars $X^I$ .  It can be expanded as
\begin{equation}
	F(X^I, \hat{A}) = \sum_{n=0}^\infty F^{(n)}(X^I)\hat{A}^n~,
\end{equation}
where each successive power of $\hat{A}$ corresponds to introducing two further derivatives to the Lagrangian, so that $F^{(n)}(X^I)$ controls the $(2+2n)$-derivative terms.  We are interested only in two-derivative and four-derivative terms, and so we can truncate this series expansion to obtain
\begin{equation}\label{eq:Fexpansion}
	F(X^I, \hat{A}) = F^{(0)}(X^I) + F^{(1)}(X^I) \hat{A}~.
\end{equation}
The new function $F^{(1)}(X^I)$ must be homogenous under rescaling of projective scalars $F^{(1)}(\lambda X^I) = F^{(1)}(X^I)$. It determines the couplings between the Weyl multiplet, vector multiplets, and chiral multiplet in the four-derivative part of the Lagrangian.  This four-derivative Lagrangian, under the truncations (\ref{eq:trunc}) and (\ref{eq:form:chiraltrunc}), is
\begin{equation}\begin{aligned}
	\mathcal{L}^{(4)} &= \frac{1}{8\pi}\bigg{[} i \mathcal{D}^\mu (F^{(1)}_I \hat{A}) \mathcal{D}_\mu \bar{X}^I + \frac{i}{4}F^{(1)}_{IJ}\mathcal{F}^{-I}_{\mu\nu}\mathcal{F}^{-\mu\nu J}\hat{A} - \frac{i}{8} F^{(1)}_I \mathcal{F}^{+I}_{\mu\nu}T^{+\mu\nu}\hat{A} \\
	&\quad - \frac{i}{32}F^{(1)} T_{\mu\nu}^+ T^{+\mu\nu}\hat{A} + \frac{i}{2}F^{(1)}_{I}\mathcal{F}^{-I}_{\mu\nu}\hat{F}^{-\mu\nu} + \frac{i}{2} F^{(1)} \hat{C} \bigg{]} +\text{h.c.}~.
\label{eq:l_4deriv}
\end{aligned}\end{equation}


\subsection{A Class of Solutions}\label{sec:class}


We are particularly interested in a class of (generally non-supersymmetric) solutions within $\mathcal{N}=2$ supergravity
determined by the two conditions:
\begin{align}
\label{eq:class:scalarsct} \partial_{\mu} X^I &= 0~,\\
\label{eq:class:curlyF0} \mathcal{F}^{+I}_{\mu\nu} &= 0~.
\end{align}
This also implies the complex conjugate equations $\partial_\mu \bar{X}^I = \mathcal{F}^{-I}_{\mu\nu}=0$.

The condition (\ref{eq:class:curlyF0}) can be re-written at two-derivative order using the definition (\ref{eq:curlyFdef}) and the auxiliary equation of motion (\ref{eq:Ttwoder}) to give
\be \left( \delta^I_K - \frac{X^I N_{KJ}X^J}{X^L N_{LM} X^M}\right)F^{+K}_{\mu\nu} =0~.\ee
For non-degenerate $N_{IJ}$, the only non-trivial solution is given by
\be \label{eq:class:EMembedding} F^{+I}_{\mu\nu} = X^I F_{\mu\nu}^+~,\ee
where at this point $F_{\mu\nu}$ is simply an arbitrary anti-symmetric two-tensor (and in particular does not yet need to satisfy a Bianchi identity). Once we also use the condition (\ref{eq:class:scalarsct}) of constant scalars, the field $F_{\mu\nu}$ becomes a genuine Maxwell field, and the resulting effective Lagrangian (at two-derivative order) following from $\mathcal{N}=2$ supergravity is simply the Einstein-Maxwell Lagrangian:
\be \mathcal{L}_{\text{eff}} = -\frac{1}{2\kappa^2}\left(R + \frac14 F_{\mu\nu}F^{\mu\nu}\right)~.\ee
For this embedding, we note that (\ref{eq:Ttwoder}) simplifies to 
\be 
\label{eq:TF}
T_{\mu\nu}^+ = 4 F_{\mu\nu}^+~,\ee
so the Weyl multiplet  ``graviphoton'' $T_{\mu\nu}^+$ is proportional to the Maxwell field $F_{\mu\nu}^+$.  Additionally, the embedding forces the $U(1)_R$ gauge field $A_\mu$ to vanish.

These Einstein-Maxwell solutions are in general not supersymmetric.  For example, general Kerr-Newman black holes will break all supersymmetries except in the non-rotating, extremal limit.  Interestingly, our Einstein-Maxwell solutions retain a remnant of the supersymmetry of the original theory: the embedding conditions (\ref{eq:class:scalarsct}) and (\ref{eq:class:curlyF0}) are exactly the conditions required for the gaugino supersymmetry variation to vanish, as discussed in~\cite{Mohaupt:2000mj,LopesCardoso:2000qm}.  We can think of non-supersymmetric Einstein-Maxwell solutions as continuous deformations of supersymmetric ones such 
that the relation between scalars and vectors demanded by the SUSY attractor mechanism is maintained. 
Then the vector multiplet fields force the gaugino variations to vanish (but do not necessarily satisfy any of the other BPS conditions).

To summarize, the conditions (\ref{eq:class:scalarsct}), (\ref{eq:class:curlyF0}) reduce the full $\mathcal{N}=2$ supergravity equations of motion to the much simpler equations of motion for Einstein-Maxwell theory.  Conversely, (\ref{eq:class:EMembedding}) defines an embedding into $\mathcal{N}=2$ supergravity of any solution to Einstein-Maxwell theory.


\section{Duality Constraints on Four-Derivative Actions}\label{sec:duality}


The Weyl anomaly (\ref{eqn:trace}) can be encoded in an effective four-derivative term in the action, as discussed in section \ref{sec:effQFT}. 
In this section we show that duality constraints on possible four-derivative terms can be quite restrictive. For Einstein-Maxwell theory, duality restricts 
the possible four-derivative terms to purely geometric curvature terms; explicit dependence on the field strength $F_{\mu\nu}$ is not possible.
This result is maintained for the embedding of Einstein-Maxwell solutions in ${\cal N}=2$ supergravity discussed in section~\ref{sec:class} but ${\cal N}=2$ supergravity
generally allows more terms.


\subsection{Einstein-Maxwell Theory and Duality}\label{sec:duality:EinsteinMaxwell}


We first review duality symmetry for Einstein-Maxwell theory and show how it restricts the possible four-derivative 
terms.

Maxwell theory (coupled minimally to gravity) has the Lagrangian
\be \label{eq:duality:EinsteinMaxwell} \mathcal{L} = -\frac{1}{2\kappa^2}\left( R + \frac14 F_{\mu\nu}F^{\mu\nu}\right)~.\ee
The dual field tensor $G_{\mu\nu}$ is defined through the relation
\be\label{eq:duality:GEM} i\tilde{G}_{\mu\nu} = 2\frac{\partial \mathcal{L}}{\partial F^{\mu\nu}}~,\ee
so that\footnote{The factor of $i$ is due to our definition (\ref{eq:N2:tF}) of the dual tensor, since then $\tilde{F}$ is purely imaginary when $F$ is real.} $G_{\mu\nu} = i\tilde{F}_{\mu\nu}$. The equations of motion and Bianchi identity can be summarized as
\be \nabla_{\mu} \left( \begin{array}{c} F\\ G\end{array}\right)^{\mu\nu} = 0~.\ee
These equations are invariant under $SO(2,\mathbb{R})$ rotations of the vector $(F_{\mu\nu},G_{\mu\nu})$ or, equivalently, $U(1)$ transformations of the (anti-)self-dual tensors $F^{\pm}_{\mu\nu}$ (defined precisely in (\ref{eq:N2:Fpm})):
\be \label{eq:duality:Maxwelldualityinf} F^{'\pm}_{\mu\nu}= e^{\pm i\varphi}F^{\pm}_{\mu\nu}~,\ee
for any phase factor $e^{\i\varphi}$. 
The $F^{\pm}_{\mu\nu}$ allow an obvious duality invariant tensor 
\be \calI_{\mu\nu\rho\sigma} \equiv F^+_{\mu\nu} F^-_{\rho\sigma}~.
\label{eqn:Itensor} \ee
All duality invariants can be formed from powers of this tensor. Lorentz invariants can then be formed by appropriate contractions of indices. 

The Einstein equations following from (\ref{eq:duality:EinsteinMaxwell}) can be written in a manifestly duality-invariant form
\be \label{eq:duality:EMEinstein} R_{\mu\nu} =  \calI_{(\mu\ \nu)\rho}^{\ \ \rho}~.\ee
The trace condition $R=0$ follows: there is no way to form a non-zero Lorentz scalar from a single $\calI_{\mu\nu\rho\sigma}$ by contracting all indices.

 In section \ref{sec:duality:invarianceL4} (and also appendix \ref{sec:appendix:dualityL4}) we will show that all four-derivative corrections corrections to the action (\ref{eq:duality:EinsteinMaxwell}) must in fact be invariant under the duality symmetry (\ref{eq:duality:Maxwelldualityinf}) even though, as is well-known, the two-derivative action (\ref{eq:duality:EinsteinMaxwell}) is \emph{not} invariant under duality (\ref{eq:duality:Maxwelldualityinf}), but rather must transform in a very particular way in order that the equations of motion respect duality symmetry \cite{Gaillard:1981rj}. In anticipation of this result we proceed to form all possible Lorentz invariants from $\calI_{\mu\nu\rho\sigma}$ and the Riemann tensor by contraction of Lorentz indices.

It is clear from the equations of motion (\ref{eq:duality:EMEinstein}) that any expression where $\calI_{\mu\nu\rho\sigma}$ appear with contracted indices reduces to the geometric invariant $R_{\mu\nu} R^{\mu\nu}$. 
There are two inequivalent ways to contract indices of two distinct $\calI_{\mu\nu\rho\sigma}$'s but one can show using the (anti-)self-duality properties of 
$F^{\pm}_{\mu\nu}$ that both also reduce to the geometric invariant $R_{\mu\nu} R^{\mu\nu}$: 
\begin{align}
\ \frac14\calI_{\mu\nu\rho\sigma} \calI^{\mu\nu\rho\sigma} & = \calI_{\mu\nu\rho\sigma} \calI^{\mu\rho\nu\sigma} = \calI_{\mu\rho\nu}^{\ \ \ \rho}\calI^{\mu\sigma\nu}_{\ \ \ \sigma}= R_{\mu\nu} R^{\mu\nu}~.
\label{eq:duality:MaxwellRic2} 
 \end{align}
 
We can also form mixed duality invariants by contracting the matter tensor $\calI_{\mu\nu\rho\sigma}$ and the Riemann tensor $R_{\mu\nu\rho\sigma}$. 
The two distinct contractions again reduce to geometric invariants
\begin{align}
\frac12\calI_{\mu\nu\rho\sigma} R^{\mu\nu\rho\sigma} &= \calI_{\mu\nu\rho\sigma} R^{\mu\rho\nu\sigma} = \calI_{\mu\rho\nu}^{\ \ \ \rho} R^{\mu\nu} = R_{\mu\nu} R^{\mu\nu}~.
 \end{align}
Thus we find that duality symmetry for Einstein-Maxwell theory restricts {\it all} the on-shell four-derivative terms to a linear combination of only ${\rm Riem}^2$ and ${\rm Ric}^2$, 
with no explicit appearance of the field strength $F_{\mu\nu}$. 

It has been noticed before that one loop corrections to Einstein-Maxwell theory reduce to pure geometry in this way \cite{Bhattacharyya:2012wz,Deser:1974cz,Deser:1975sx} and a relation to duality was mentioned \cite{Deser:1975sx} but we are not aware of a detailed exposition of this feature. 


\subsection{Symplectic Duality Symmetry}\label{sec:N2duality}


We want to discuss four-derivative duality invariants in a much more general theory of ${\cal N}=2$ supergravity. To get started, we first review the extended symplectic duality 
symmetry of $\mathcal{N}=2$ supergravity.

In a theory with $n_V+1$ $U(1)$ gauge fields (and no explicit sources for the gauge fields) there is a $U(n_V+1)$ compact duality symmetry that rotates the gauge fields and their dual tensors into each other. When there are also scalars in the theory that transform under duality, such as in $\mathcal{N}=2$ supergravity, the duality symmetry can further be extended to a non-compact (sub)group of $Sp(2n_V+2,\mathbb{R})$ \cite{Gaillard:1981rj}.\footnote{Symplectic duality for $\mathcal{N}=2$ is discussed in detail in e.g. \cite{deWit:1996gjy,deWit:1997ad,deWit:2001pz}, and also reviewed in e.g. \cite{Mohaupt:2000mj,Freedman:2012zz}.}

The dual field strengths $G_{I\mu\nu}$ (with $I=0,\ldots,n_V$) generalizing (\ref{eq:duality:GEM}) are
\be i\tilde{G}_{I\mu\nu} = \frac{\partial (8\pi\mathcal{L})}{\partial F^{I\mu\nu}} ~.\ee
In the case of the on-shell two-derivative Lagrangian (\ref{eq:lgonshell})
\be \label{eq:duality:Gonshell} G_{I\, \mu\nu}^+ = \mathcal{N}_{IJ}F^{+J}_{\mu\nu} ~.\ee
Under the $Sp(2n_V+2,\mathbb{R})$ symplectic duality symmetry of $\mathcal{N}=2$ supergravity, the field strengths $F^I_{\mu\nu}$ and the dual field strengths $G_{I\mu\nu}$ form a symplectic vector
\be \label{eq:duality:bF} \mathbb{F}_{\mu\nu}\equiv(F^I_{\mu\nu},G_{I\, \mu\nu}) ~,\ee
that transforms under duality as
\be \left( \begin{array}{c} F^I_{\mu\nu}\\ G_{J\, 	\mu\nu} \end{array}\right)\rightarrow \left(\begin{array}{cc} U^I_{\ K} & Z^{IL}\\ W_{JK} & V_J^{\ L}\end{array}\right) \left( \begin{array}{c} F^K_{\mu\nu}\\ G_{L\, 	\mu\nu} \end{array}\right)~,\ee
with $U,Z,W,V$ real matrices satisfying $U^T W-W^T U=Z^TV-V^TZ=0$ and $U^TV-W^TZ=\mathbb{I}$. The infinitesimal version of this transformation is
\be \label{eq:duality:FGsymplinf} \delta \left( \begin{array}{c} F^I_{\mu\nu}\\ G_{J\, \mu\nu} \end{array}\right) = \left(\begin{array}{cc} A^I_{\ K} & B^{IL}\\ C_{JK} & (-A^T)_J^{\ L}\end{array}\right) \left( \begin{array}{c} F^K_{\mu\nu}\\ G_{L\, 	\mu\nu} \end{array}\right)~,\ee
where $B$ and $C$ are symmetric (real) matrices and $A$ is arbitrary (real). The vector multiplet scalars $X^I$ also transform under these transformations, forming a symplectic vector with the prepotential derivatives $F_I$ as
\be \label{eq:duality:bX} \mathbb{X}\equiv (X^I,F_I) ~.\ee
We can form symplectic scalars by taking the symplectic product of any two symplectic vectors $\mathbb{A},\mathbb{B}$
\be \mathbb{A}\cdot \mathbb{B} \equiv \mathbb{A}\, \Omega\,  \mathbb{B},\qquad \Omega \equiv \left(\begin{array}{cc} 0& \mathbb{I}\\ -\mathbb{I} & 0\end{array}\right)~.\ee
Such symplectic scalars generalize the invariant tensor (\ref{eqn:Itensor}) from Einstein-Maxwell theory. As in that example, they generally transform under Lorentz symmetry (the Lorentz indices may be uncontracted at this point). 

The prepotential $F(X^I)$ is not a symplectic scalar even though it has no symplectic index $I$. By integrating how the functions $F_I(X^J)$ change under duality transformations, one can find how the prepotential $F(X^I)$ transforms. The result is that, for a given prepotential $F(X^I)$, the symplectic transformations (\ref{eq:duality:FGsymplinf}) fall into two categories: the transformations that leave the prepotential \emph{invariant}, i.e. transformations that preserve the functional form of $F(X^I)$: $F(X^I)\rightarrow F(X^I+\delta X^I)$; and the transformations that change the functional form of the prepotential. The former transformations are true \emph{symmetries} of the particular $\mathcal{N}=2$ theory with given $F(X^I)$, while the latter transformations are not symmetries but rather \emph{symplectic reparametrizations} that transform the equations of motion of the theory into \emph{equivalent} but different equations of motion \cite{deWit:2001pz,Mohaupt:2000mj}. 

The generalized prepotential $F(X^I,\hat{A})$ needed to introduce four-derivative terms depends on a (duality-invariant) chiral scalar $\hat{A}$ with 
Weyl weight two. In this setting we can form the partial derivative $F_A$ that also has no symplectic index $I$. This derivative has zero Weyl weight and
it is always a symplectic scalar \cite{deWit:1996gjy,Mohaupt:2000mj}. This will be important in the discussion later on, particularly in sections \ref{sec:duality:invarianceL4} and \ref{sec:susy:invariant}.


\subsection{Duality (In)variance of Four-Derivative Corrections}\label{sec:duality:invarianceL4}


It is well-known that the (two-derivative) Lagrangian of a theory with duality symmetry is not itself invariant under duality transformations; the symmetry is manifest only at the level of the equations of motion. The transformation properties of four-derivative terms under duality symmetry are less familiar. We will show that the (on-shell) four-derivative corrections must be duality invariant already at the Lagrangian level in the situations that we are most interested in, but not in general. 

Our claim generalizes a result by Gaillard and Zumino \cite{Gaillard:1981rj}. They showed that, if a Lagrangian $\mathcal{L}$ depends on a duality-invariant parameter $\alpha$, and further that the duality transformations of the scalars do not depend on $\alpha$, then $\partial_{\alpha}\mathcal{L}$ is duality invariant.
We can apply this argument by identifying $\alpha$ as the coupling constant $c_4$ introduced in section 2 as the coefficient that multiplies the four-derivative action 
\be \mathcal{L}^{(4)} = c_4 \hat{\mathcal{L}}^{(4)}~,\ee
and conclude that the four-derivative action $\hat{\mathcal{L}}^{(4)}$ is duality invariant. This reasoning certainly applies when there are no scalars at all, such as the simple Einstein-Maxwell theory. Therefore the four-derivative corrections must be duality invariant in this case, justifying the assumption made in section \ref{sec:duality:EinsteinMaxwell}.

However, for higher-derivative corrections to ${\cal N}=2$ supergravity we can generally \emph{not} choose an $\alpha$ that the scalars do not depend on: duality acts on coupling constants and, specifically, the couplings encoded in the prepotential (\ref{eq:Fexpansion}) are not duality invariant. Generalizing the result of Gaillard and Zumino  \cite{Gaillard:1981rj} to take dependence of the scalar transformations on the coupling constant $\alpha$ into account we find (through a simple calculation spelled out in appendix \ref{sec:appendix:transfL4}) 
\be\label{eq:duality:transfL4} \delta \mathcal{L}^{(4)} = -B^{IJ}F_J^{(1)} \frac{\partial \mathcal{L}^{(2)}}{\partial X^I} -  B^{IJ}F_{JK}^{(1)}\partial_{\mu}X^K \frac{\partial \mathcal{L}^{(2)}}{\partial(\partial_{\mu}X^I)}~. \ee
Thus, in general, the four-derivative corrections to $\mathcal{N}=2$ supergravity are \emph{not} duality invariant. Fortunately, they are not arbitrary: the transformation properties of the four-derivative Lagrangian $\mathcal{L}^{(4)}$ are completely determined by the two-derivative Lagrangian $\mathcal{L}^{(2)}$.

We are particulatly interested in the class of solutions introduced in section \ref{sec:class} where the scalars are constant and the superconformal field strength $\mathcal{F}^+_{\mu\nu}$ vanishes. In this case the expression (\ref{eq:l_2deriv}) of the (off-shell) two-derivative Lagrangian $\mathcal{L}^{(2)}$ gives
\be \left[\frac{\partial \mathcal{L}^{(2)}}{\partial(\partial_{\mu}X^I)}\right]_{\partial_{\mu}X^I=0}=0~,\ee
and,  remembering the dependence of $\mathcal{F}_{\mu\nu}^+$ on the scalars $X^I$, we have
\be \left[\frac{\partial \mathcal{L}^{(2)}}{\partial X^I}\right]_{\partial_{\mu}X^I=0,\, \mathcal{F}^+_{\mu\nu}=0} = \left[\frac{\partial \left(-\frac{i}{8} F_I^{(0)} \mathcal{F}^{+I}_{\mu\nu}T^{+\mu\nu} -\frac{i}{32} F^{(0)} (T^+)^2 + \text{h.c.}\right)}{\partial X^I}\right]_{\mathcal{F}^+_{\mu\nu}=0} = 0~, \ee
identically (without using the equation of motion). We conclude that the four-derivative Lagrangian must be duality invariant when the scalars are constant and $\mathcal{F}^+_{\mu\nu}$ vanishes
\be \left[\delta \mathcal{L}^{(4)}\right]_{\partial_{\mu}X^I=0,\ \mathcal{F}^+_{\mu\nu}=0} = 0~.\ee
This result does not in any way depend on supersymmetry, neither of the theory nor of the solution. 

In section \ref{sec:susy:invariant} below we show that in ${\cal N}=2$ supergravity the four-derivative corrections are given by (\ref{eq:susy:L4simpl}), an expression that only depends on the symplectic scalar function $F_A=F^{(1)}$ and the (symplectically invariant) components of the Weyl multiplet. The discussion in this section shows that corrections of this form must be invariant under duality at the level of the Lagrangian. 


\subsection{Four-Derivative Symplectic Invariants with Constant Scalars \label{sec:duality:fourder}}



It is interesting to investigate how much we can constrain four-derivative terms using symplectic duality invariance alone. In this section,  we restrict ourselves to the case with constant scalars and set all fermions to zero but impose no other restrictions on the bosonic fields. This is a generalization of the discussion at the end of section \ref{sec:duality:EinsteinMaxwell} where we showed that four-derivative corrections to Einstein-Maxwell theory can always be written in terms of curvature invariants involving only geometry.\footnote{We are not yet setting $\mathcal{F}^{+I}_{\mu\nu}=0$ so, as explained in section \ref{sec:duality:invarianceL4}, the full four-derivative Lagrangian will generally \emph{not} be a duality-invariant. The duality-invariants we find in this section should therefore be viewed as (at most) \emph{part} of the four-derivative Lagrangian for constant scalars when $\mathcal{F}^{+I}_{\mu\nu}\neq 0$; another (mandatory) part of the Lagrangian must be given by a non-duality-invariant term that transforms according to (\ref{eq:duality:transfL4}).}

As a first step we classify all invariants under duality we can construct using at most two symplectic vectors and at most 
four covariant derivatives. We do not yet impose Lorentz invariance. We will use the notation we introduced in (\ref{eq:duality:bF}) and (\ref{eq:duality:bX}) for the symplectic vector of the (anti-)self-dual field strengths $\bF^{\pm}_{\mu\nu}$ and the scalar symplectic vector $\bX$ (and its complex conjugate $\bbX$). Our starting point is the on-shell Lagrangian (\ref{eq:lgonshell}) where auxiliary fields have been integrated out.  

At zero-derivative order, the only symplectic invariants are $\bX\cdot \bX=0$ and $\bX\cdot \bbX=8\pi i/\kappa^2$ (where we have used (\ref{eq:trunc})). At one-derivative order, we have the symplectic invariant
\be\label{eq:duality:ctescalarsT} T^+_{\mu\nu} = -\frac{i}{2\pi}\kappa^2 \bF^+_{\mu\nu}\cdot \bbX~, \ee
and its complex conjugate, where we have recognized the auxiliary Weyl multiplet field $T^+_{\mu\nu}$ from its two-derivative equation of motion (\ref{eq:Ttwoder}). The only other possible symplectic invariant with one derivative is $\bF^+_{\mu\nu}\cdot \bX$, which vanishes 
\be \bF^+_{\mu\nu}\cdot \bX = F^{+I}_{\mu\nu} F_I - G_{I\, \mu\nu}^+ X^I =  F^{+I}_{\mu\nu} F_I - \mathcal{N}_{IJ} F^{+J}_{\mu\nu} X^I =  F^{+I}_{\mu\nu} F_I -  F^{+J}_{\mu\nu} F_J = 0~,\ee
using the explicit form (\ref{eq:duality:Gonshell}) for the dual tensor $G^+_{I\, \mu\nu}$ and the special geometry identity $\mathcal{N}_{IJ}X^I = F_I$.

At two-derivative order, we have the symplectic invariants
\be \calI_{2\mu\nu\rho\sigma}= \bF^+_{\mu\nu}\cdot \bF^-_{\rho\sigma}, \qquad \nabla_{\rho} T_{\mu\nu}^+ = -\frac{i}{2\pi}\kappa^2 \nabla_{\rho} \bF^+_{\mu\nu}\cdot \bbX, \qquad R_{\mu\nu\rho\sigma}~.\ee
There are two other possible candidates but both vanish identically $\bF^+_{\mu\nu}\cdot \bF^+_{\rho\sigma}=\nabla_\rho(\bF^+_{\mu\nu}\cdot \bX)=0$
using (\ref{eq:duality:Gonshell}). Using similar arguments, at three-derivative order we can have
\be \nabla_{\lambda} \bF^+_{\mu\nu}\cdot \bF^+_{\rho\sigma},\qquad \nabla_{\lambda} \bF^+_{\mu\nu}\cdot \bF^-_{\rho\sigma},\qquad \nabla_\rho\nabla_\sigma \bF^+_{\mu\nu}\cdot \bbX~,\ee
and their complex conjugates. Finally, at four-derivative order, we can have
\be \nabla_\lambda \bF^+_{\mu\nu}\cdot \nabla_\omega \bF^+_{\rho\sigma}, \qquad \nabla_\lambda\nabla_\omega\bF^+_{\mu\nu}\cdot \bF^-_{\rho\sigma}, \qquad \nabla_\lambda\bF^+_{\mu\nu}\cdot \nabla_\omega \bF^-_{\rho\sigma}, \qquad \nabla_\lambda\nabla_\omega\bF^+_{\mu\nu}\cdot  \bF^-_{\rho\sigma}~,\ee
and their complex conjugates.

Having now determined all possible symplectic invariants with at most two symplectic vectors, the next step is to multiply such invariants together and contract Lorentz indices to form four-derivative terms that are invariant under Lorentz symmetry as well as symplectic invariance. There are numerous options but the physically interesting ones are subject to further constraints: 
\begin{itemize}
\item
Candidate terms for four-derivative corrections to $\mathcal{N}=2$ supergravity must have vanishing $U(1)_R$ charge. This is restrictive since $\bX$ is charged under 
$U(1)_R$;
\item
We can use the two-derivative on-shell Einstein equation to trade $\bF^+_{\mu\rho}\cdot \bF^{-\ \rho}_{\nu}$ for $R_{\mu\nu}$ (a generalization of (\ref{eq:duality:EMEinstein})); \item
We can discard terms that are equivalent up to a total derivative. 
\end{itemize}
Using all of these properties we find (through straightforward but tedious calculations involving the (anti-)self-duality of $\bF^{\pm}_{\mu\nu}$) that there are exactly five independent four-derivative symplectic invariant terms:
\be \label{eq:duality:fourderinvfinal} R_{\mu\nu\rho\sigma}R^{\mu\nu\rho\sigma}~,~ R_{\mu\nu}R^{\mu\nu}~,~ \nabla_{\mu} T^{+\mu\nu} \nabla_{\rho} T^{-\rho}_{\ \ \ \nu}~,~ R_{\mu\nu} T^{+\mu}_{\ \ \  \rho}T^{-\nu\rho}~,~ T^{-\ \rho}_{\ \mu} T^{-\mu\nu}T^{+\ \sigma}_{\ \nu}T^{+}_{\rho\sigma} ~.\ee
We spell out more details of the calculation leading to (\ref{eq:duality:fourderinvfinal}) in appendix \ref{sec:appendix:invariants}.

It is interesting that, even with the minimal assumptions made in this subsection, all these terms involve only fields from the Weyl multiplet; all explicit dependence on the vector multiplets has been eliminated using symmetries and equations of motion.


\subsection{The Einstein-Maxwell Embedding in $\mathcal{N}=2$ Supergravity}

We are particularly interested in the embedding of Einstein-Maxwell theory in $\mathcal{N}=2$ supergravity with any number $n_V$ of $\mathcal{N}=2$ vector multiplets. 
As discussed in section \ref{sec:class}, the embedding presumes scalars $X^I={\rm constant}$ fixed at any constant value and, given those scalars, specifies the $\mathcal{N}=2$ vector fields as (\ref{eq:class:EMembedding}) 
\be \label{eq:duality:EMembedding} F^{+I}_{\mu\nu} = X^I F^+_{\mu\nu}~,\ee
for some Maxwell gauge field $F_{\mu\nu}$. Since this setting has constant scalars the results from the previous subsection applies. However, in addition, the 
Einstein-Maxwell embedding (\ref{eq:duality:EMembedding}) demands that the superconformal curvature vanishes $\mathcal{F}^{+I}_{\mu\nu} = 0$. In this setting 
the antisymmetric tensor $T^+_{\mu\nu}$ (\ref{eq:duality:ctescalarsT}) in the Weyl multiplet reduces to the Einstein-Maxwell field strength $F^+_{\mu\nu}$, as noted 
in (\ref{eq:TF}). Then 
the four-derivative invariants in (\ref{eq:duality:fourderinvfinal}) either vanish due to the Maxwell equations/Bianchi identity on the Einstein-Maxwell field strength $F_{\mu\nu}$
or reduces, through the Einstein equation (\ref{eq:duality:EMEinstein}) for Einstein-Maxwell theory, to pure geometry. Thus the four-derivative invariants respect the duality symmetry of the Maxwell theory defined by $F_{\mu\nu}$ discussed in \ref{sec:duality:EinsteinMaxwell}, and so we are left with the two independent invariants $W^2$ and $E_4$. 

It is interesting to trace the origin of the Maxwell duality symmetry of $F_{\mu\nu}$ in the underlying $\mathcal{N}=2$ supergravity theory. Indeed, at first sight this duality symmetry is quite mysterious: since $F_{\mu\nu}^{+I}$ transforms like $X^I$ under the $\mathcal{N}=2$ symplectic duality transformations, the embedded Maxwell field $F^+_{\mu\nu}$ defined in (\ref{eq:duality:EMembedding}) is {\it invariant} under $\mathcal{N}=2$ symplectic duality. Therefore the duality symmetry of the Maxwell theory is \emph{not} a subset of the $\mathcal{N}=2$ duality symmetry.
 
We must instead pay attention to the $U(1)_R$ symmetry of $\mathcal{N}=2$ supergravity: $F^{+I}_{\mu\nu}$ is uncharged and $X^I$ is charged so, according to (\ref{eq:duality:EMembedding}), the embedded Maxwell field $F^+_{\mu\nu}$ must be charged under the $U(1)_R$ symmetry (with charge opposite to that of $X^I$).  
Therefore, it transforms as (\ref{eq:duality:Maxwelldualityinf}) under the global 
$U(1)_R$ and we conclude that \emph{the $U(1)$ duality symmetry of the embedded Maxwell field $F_{\mu\nu}$ is identified with the $U(1)_R$ global symmetry of the $\mathcal{N}=2$ theory.} Thus it is ultimately the $U(1)_R$ of the underlying $\mathcal{N}=2$ supergravity that is responsible for the only allowed (on-shell) four-derivative invariants being the geometric curvature invariants Riem$^2$ and Ric$^2$ or, equivalently, $W^2$ and $E_4$.


\section{Supersymmetry and $c=0$}
\label{sec:resultssusy}


In section~\ref{sec:duality} we showed that the only four-derivative terms allowed in our Einstein-Maxwell embedding (introduced in section~\ref{sec:class})
are the geometric invariants $E_4$ and $W^2$.  Duality prevents explicit dependence on matter. In this section, we show how supersymmetry 
further constrains the four-derivative terms such that only the Euler invariant $E_4$ can appear, hence proving that the $c$-anomaly vanishes.

This section proceeds as follows. In section~\ref{sec:susy:invariant} we discuss simplifications of the four-derivative Lagrangian (\ref{eq:l_4deriv}) due to the form of
the Einstein-Maxwell embedding introduced in section~\ref{sec:class} .  We will go on to discuss the two known four-derivative chiral multiplets, the $\mathbf{W}^2$ multiplet and the $\mathbb{T}(\log\bar{\mathbf{\Phi}})$ multiplet, in section~\ref{sec:susy:chirals}.  We will use use the details of these chiral multiplets to show how we are forced to have $c=0$ in section~\ref{sec:susy:c0}.


\subsection{Four-Derivative Action in the Einstein-Maxwell Embedding}
\label{sec:susy:invariant}


The general form of the four-derivative part of the Lagrangian is given in (\ref{eq:l_4deriv}).  In the Einstein-Maxwell embedding (\ref{eq:class:EMembedding})
we set
\begin{equation}
	F_{\mu\nu}^{+I} = X^I F^+_{\mu\nu}~, \quad \partial_\mu X^I = 0~, 
\end{equation}
so the supercovariant field strengths $\mathcal{F}^{\pm I}_{\mu\nu}$ vanish, and then the Lagrangian simplifies to
\begin{equation}\label{eq:susy:L4simpl}
	\mathcal{L}^{(4)} = \frac{i}{16\pi}F^{(1)}(X^I)\left(\hat{C} - \frac{1}{16}T^+_{\mu\nu}T^{+\mu\nu}\hat{A}\right) + \text{h.c.}~.
\end{equation}
We recall from (\ref{eq:Fexpansion}) that the four-derivative prepotential term $F_A = F^{(1)}$ is a function of the vector multiplet scalars, which are all set to a constant in the Einstein-Maxwell embedding.  In this context the four-derivative Lagrangian is therefore given by the supersymmetric invariant
\begin{equation}
	\mathcal{L}^-_{\hat{\mathbf{A}}} =\frac{1}{64}\left(\hat{C} - \frac{1}{16}T^+_{\mu\nu}T^{+\mu\nu}\hat{A}\right)~,
\label{eq:susy:chiraldensity}
\end{equation}
plus its Hermitian conjugate. This shows that, when considering the class of Einstein-Maxwell solutions discussed in section~\ref{sec:class}, the only four-derivative 
Lagrangian that respects supersymmetry is made up entirely of Weyl and chiral multiplet fields; no couplings between the chiral and vector multiplets are allowed when the supercovariant field strengths vanish and the scalars are constant.

This supersymmetric invariant (\ref{eq:susy:chiraldensity}) matches the chiral multiplet density formula discussed in~\cite{deWit:2010za}, after the truncation (\ref{eq:form:chiraltrunc}) has been imposed on the chiral multiplet field content.


\subsection{Chiral Multiplet Supersymmetric Invariants}
\label{sec:susy:chirals}


As discussed in section~\ref{sec:form:fourderiv}, our interest in the chiral multiplet $\hat{\mathbf{A}}$ is to introduce higher-derivative terms into the action.  In this context the fields that make up the chiral multiplet are not independent fields, but rather composites of fields that are already introduced as components of other superfields. In order to introduce four-derivative interactions (e.g. $R^2$, $F^4$, etc.) the chiral multiplet must have a Weyl weight $w=2$. This also guarantees that the supersymmetric invariant (\ref{eq:susy:chiraldensity}) is both symplectically invariant and $U(1)_R$ invariant, as required by the discussion in section~\ref{sec:duality:fourder}.  

The only known chiral multiplets that fit these criteria are the $\mathbf{W}^2$ multiplet (introduced in~\cite{Bergshoeff:1980is} and reviewed in detail in \cite{Mohaupt:2000mj,LopesCardoso:2000qm,LopesCardoso:1998tkj}) and 
the $\mathbb{T}(\log\bar{\mathbf{\Phi}})$ multiplet (introduced in~\cite{Butter:2013lta}).  In the following, we will discuss the basic structures needed to establish the form of the supersymmetric invariant (\ref{eq:susy:chiraldensity}) for each of these multiplets. These multiplets are discussed in more detail in appendix~\ref{sec:appN2}. 

We first discuss $\mathbf{W}^2$, the more familiar of the two. Constraints can be imposed on the Weyl multiplet (\ref{eqn:Weylmultiplet}) such that it forms a reduced chiral multiplet, denoted by $\mathbf{W}_{ab}$.  Using
standard rules for performing algebraic operations on chiral multiplets, we can take the product of this reduced chiral multiplet with itself to obtain a new chiral multiplet
$\mathbf{W}^2 \equiv	\mathbf{W}_{ab} \mathbf{W}^{ab}$ that is a Lorentz scalar.  The components of $\mathbf{W}^2$ are given in appendix~\ref{sec:appN2:higher}. 

The supersymmetrized invariant (\ref{eq:susy:chiraldensity}) with chiral field $\hat{\mathbf{A}} = \mathbf{W}^2$ is
\begin{equation}\begin{aligned}
	\mathcal{L}^-_{\mathbf{W}^2} &= \frac{1}{64}\left({C}|_{\mathbf{W}^2} - \frac{1}{16} T^{+}_{\mu\nu}T^{+\mu\nu} {A}|_{\mathbf{W}^2}\right) \\
	&= \frac{1}{2}W_{\mu\nu\rho\sigma}W^{\mu\nu\rho\sigma} + \frac{i}{2}{}^*W_{\mu\nu\rho\sigma}W^{\mu\nu\rho\sigma} + \frac{1}{8}R^\mu_{~\nu}T^-_{\mu\rho}T^{+\nu\rho} + 3 D^2 \\
	&\quad + \frac{1}{1024}T^-_{\mu\nu}T^{-\mu\nu}T^+_{\rho\sigma}T^{+\rho\sigma} - \frac{1}{4}T^{-\mu\nu}\mathcal{D}_\mu \mathcal{D}^\rho T^+_{\rho\nu}  \\
	&\quad  - 2A^-_{\mu\nu}A^{-\mu\nu} + \frac{1}{2} \mathcal{V}_{\mu\nu~j}^{-~i} \mathcal{V}^{-\mu\nu j}_{~~~~~i} + \text{(fermions)} ~.
\label{eq:superweyl1}
\end{aligned}\end{equation}
$\mathcal{L}^-_{\mathbf{W}^2}$ thus contains a $W_{\mu\nu\rho\sigma}W^{\mu\nu\rho\sigma}$ term, in addition to many other terms formed from Weyl multiplet fields.  
It is the supersymmetric completion of $W_{\mu\nu\rho\sigma}W^{\mu\nu\rho\sigma}$ denoted schematically in (\ref{eqn:fourderivsusy}) as ``$W^2 + \text{SUSY matter}$".

Next, we discuss the less familiar $\mathbb{T}(\log\bar{\mathbf{\Phi}})$ multiplet. For an arbitrary chiral multiplet $\mathbf{\Phi}$, we can take its Hermitian conjugate and then (using chiral multiplet algebra rules) take the logarithm of this Hermitian conjugate, resulting in the anti-chiral multiplet $\log\bar{\mathbf{\Phi}}$ with Weyl weight $w = 0$.  We can act on this multiplet with the kinetic operator $\mathbb{T}$, which introduces two powers of derivatives in order to make the multiplet kinetic~\cite{deWit:1980lyi}.  This new kinetic chiral multiplet has Weyl weight $w = 2$ and is denoted $\mathbb{T}(\log\bar{\mathbf{\Phi}})$~\cite{Butter:2013lta}.  The field content of the $\mathbb{T}(\log\bar{\mathbf{\Phi}})$ multiplet are discussed in appendix~\ref{sec:appN2:higher}.

As discussed in~\cite{Butter:2013lta}, the supersymmetrized invariant (\ref{eq:susy:chiraldensity}) derived from the chiral multiplet $\hat{\mathbf{A}} = \mathbb{T}(\log\bar{\mathbf{\Phi}})$ can be written as
\begin{equation}\begin{aligned}
	\mathcal{L}^-_{\mathbb{T}(\log\bar{\mathbf{\Phi}})} &= \frac{1}{64}\left(C|_{\mathbb{T}(\log\bar{\mathbf{\Phi}})} - \frac{1}{16} T^{+}_{\mu\nu}T^{+\mu\nu} A|_{\mathbb{T}(\log\bar{\mathbf{\Phi}})}\right) \\
	&= - R_{\mu\nu}R^{\mu\nu} + \frac{1}{3}R^2 - 3 D^2 - \frac{1}{8}R^\mu_{~\nu}T^-_{\mu\rho}T^{+\nu\rho} - \frac{1}{1024}T^-_{\mu\nu}T^{-\mu\nu}T^+_{\rho\sigma}T^{+\rho\sigma} \\
	&\quad+ \frac{1}{4}T^{-\mu\nu}\mathcal{D}_\mu \mathcal{D}^\rho T^+_{\rho\nu} +  A_{\mu\nu}A^{\mu\nu} - \frac{1}{2}\mathcal{V}_{\mu\nu~j}^{+~i} \mathcal{V}^{+\mu\nu j}_{~~~~~i} \\
	&\quad+ \frac{1}{2w}\nabla_a \mathbf{V}^a + \text{(fermions)}~,
\label{eq:susy:tlogmultiplet}
\end{aligned}\end{equation}
where $\mathbf{V}^a$ is given in terms of $\bar{A}|_{\log\bar{\mathbf{\Phi}}}$, $F^+_{ab}|_{\log\bar{\mathbf{\Phi}}}$, and the Weyl multiplet fields as
\begin{equation}\begin{aligned}
	\mathbf{V}^a &= 4 \mathcal{D}^a \mathcal{D}^2 \bar{A}|_{\log\bar{\mathbf{\Phi}}} - 8 R^{ab}\mathcal{D}_b \bar{A}|_{\log\bar{\mathbf{\Phi}}} + \frac{8}{3}R \mathcal{D}^a \bar{A}|_{\log\bar{\mathbf{\Phi}}} - 8i A^{ab} \mathcal{D}_b \bar{A}|_{\log\bar{\mathbf{\Phi}}} \\
	&\quad - T^{-ac}T^+_{bc}\mathcal{D}^b \bar{A}|_{\log\bar{\mathbf{\Phi}}} + \frac{1}{2}(\mathcal{D}^a T^+_{bc})F^{+bc}|_{\log\bar{\mathbf{\Phi}}} + 4T^{+ac}\mathcal{D}^b F^+_{bc}|_{\log\bar{\mathbf{\Phi}}} \\
	&\quad + w\bigg{[}\frac{2}{3}\mathcal{D}^a R - 4 \mathcal{D}^a D - \frac{1}{2}\mathcal{D}^b(T^{-ac}T^+_{bc})\bigg{]}~,
\label{eq:susy:va}
\end{aligned}\end{equation}
and $w$ is the Weyl weight of the chiral multiplet $\mathbf{\Phi}$.  It is important to note that the only dependence in (\ref{eq:susy:tlogmultiplet}) on the details of the chiral 
multiplet $\mathbf{\Phi}$ is in $\mathbf{V}^a$.  This means that, no matter what $\mathbf{\Phi}$ is taken as starting point for the construction, the resulting supersymmetric invariants are the same up to a total derivative.

The supersymmetric completion of the Euler invariant (denoted schematically as ``$E_4 + \text{SUSY matter}$" in (\ref{eqn:fourderivsusy})) is the sum of the two 
four derivative terms introduced in this subsection:
\begin{equation}\begin{aligned}
	\mathcal{L}_{\chi}^- &= \mathcal{L}_{\mathbf{W}^2}^- + \mathcal{L}^-_{\mathbb{T}(\log\bar{\mathbf{\Phi}})} \\
	&= \frac{1}{2}E_4 + \frac{i}{2} {}^*W_{\mu\nu\rho\sigma}W^{\mu\nu\rho\sigma} + A_{\mu\nu}\tilde{A}^{\mu\nu} + \frac{1}{2}\mathcal{V}_{\mu\nu~j}^{~~i}\tilde{\mathcal{V}}^{\mu\nu j}_{~~~i} + \frac{1}{2 w}\nabla_a \mathbf{V}^a + \text{(fermions)}~.
\end{aligned}\end{equation}

As discussed in section~\ref{sec:form:field}, we can consistently truncate the full off-shell supergravity theory down to one with only a subset of the full bosonic content by using the truncation ansatz (\ref{eq:trunc}).  The result of this truncation,  when applied to the supersymmetric invariants, is
\begin{equation}\begin{aligned}
	\mathcal{L}^-_{\mathbf{W}^2} &= \frac{1}{2}E_4 + \frac{i}{2}{}^*W_{\mu\nu\rho\sigma}W^{\mu\nu\rho\sigma} + \left(R_{\mu\nu} + \frac{1}{16}T^-_{\mu\rho}T^{+\rho}_\nu\right)^2 - \frac{1}{4}T^{-\mu\nu}\mathcal{D}_\mu \mathcal{D}^\rho T^+_{\rho\nu}  - 2A^-_{\mu\nu}A^{-\mu\nu}~, \\
	\mathcal{L}^-_{\mathbb{T}(\log\bar{\mathbf{\Phi}})} &= - \left(R_{\mu\nu} + \frac{1}{16}T^-_{\mu\rho}T^{+\rho}_\nu\right)^2 + \frac{1}{4}T^{-\mu\nu}\mathcal{D}_\mu \mathcal{D}^\rho T^+_{\rho\nu} +  A_{\mu\nu}A^{\mu\nu} + \frac{1}{2w}\nabla_a \mathbf{V}^a~, \\
	\mathcal{L}^-_{\chi} &= \frac{1}{2}E_4 + \frac{i}{2} {}^*W_{\mu\nu\rho\sigma}W^{\mu\nu\rho\sigma} + A_{\mu\nu}\tilde{A}^{\mu\nu} + \frac{1}{2 w}\nabla_a \mathbf{V}^a~.
\label{eq:susy:invariants}
\end{aligned}\end{equation}


\subsection{Supersymmetric Invariants in the Einstein-Maxwell Embedding}
\label{sec:susy:c0}


The final equations for the supersymmetric invariants (\ref{eq:susy:invariants}) are for any solution that satisfies the consistent truncation (\ref{eq:trunc}).  
 We now further restrict to Einstein-Maxwell solutions that result from the Einstein-Maxwell embedding (\ref{eq:class:EMembedding}). 
Then the remaining auxiliary fields are set to
\begin{equation}
	T^+_{\mu\nu} = 4 F^+_{\mu\nu}~, \quad A_\mu = 0~,
\end{equation}
where $F_{\mu\nu}$ is a $U(1)$ field strength that sources the geometry via an effective Einstein-Maxwell action
\begin{equation}
	\mathcal{L}_\text{eff} = -\frac{1}{2\kappa^2}\left(R + \frac{1}{4}F_{\mu\nu}F^{\mu\nu}\right)~.
\label{eq:susy:effmax}
\end{equation}
For such backgrounds the supersymmetric invariants (\ref{eq:susy:invariants}) simplify to
\begin{equation}\begin{aligned}
	\mathcal{L}^-_{\mathbf{W}^2} &= \frac{1}{2}E_4 + \frac{i}{2}{}^*W_{\mu\nu\rho\sigma}W^{\mu\nu\rho\sigma} + \left(R_{\mu\nu} + F^-_{\mu\rho}F^{+\rho}_\nu\right)^2 - \frac{1}{4}F^{-\mu\nu}\nabla_\mu \nabla^\rho F^+_{\rho\nu}~, \\
	\mathcal{L}^-_{\mathbb{T}(\log\bar{\mathbf{\Phi}})} &= - \left(R_{\mu\nu} + F^-_{\mu\rho}F^{+\rho}_\nu\right)^2 + \frac{1}{4}F^{-\mu\nu}\nabla_\mu \nabla^\rho F^+_{\rho\nu} + \frac{1}{2w}\nabla_a \mathbf{V}^a~, \\
	\mathcal{L}^-_{\chi} &= \frac{1}{2}E_4 + \frac{i}{2} {}^*W_{\mu\nu\rho\sigma}W^{\mu\nu\rho\sigma} + \frac{1}{2 w}\nabla_a \mathbf{V}^a~.
\label{eq:susy:invariants2}
\end{aligned}\end{equation}

The Einstein equation (\ref{eq:twoderEinstein}) and the Maxwell-Bianchi equations (\ref{eq:twoderFEOM}), (\ref{eq:twoderFBI}) for Einstein-Maxwell embedding solutions become
\begin{equation}
	R_{\mu\nu} = -F^-_{\mu\rho}F^{+\rho}_\nu~, \quad \nabla_\mu F^{\pm\mu\nu} = 0~,
\label{eq:susy:eom}
\end{equation}
which are just the familiar equations of motion for the effective action (\ref{eq:susy:effmax}).  If we now take the allowed four-derivative Lagrangians in (\ref{eq:susy:invariants2}), put them on-shell by using these Einstein-Maxwell equations of motion (\ref{eq:susy:eom}), and drop any total derivative terms in the Lagrangians\footnote{Although the Euler invariant $E_4$ can be written as a total derivative in four dimensions, it is a total derivative acting on (non-covariant) Christoffel symbols that do not fall off to zero at infinity.  Its contribution to the anomalous trace of the stress tensor in (\ref{eqn:trace}) is therefore not automatically zero and is instead proportional to the Euler characteristic (\ref{eqn:eulerinvariant}) of the spacetime.  The total derivative terms we drop are ${}^*W_{\mu\nu\rho\sigma}W^{\mu\nu\rho\sigma}$ and $\nabla_a \mathbf{V}^a$, both of which give a vanishing contribution to (\ref{eqn:trace}).}, we find that they collapse almost entirely:
\begin{equation}
	\mathcal{L}_{\mathbf{W}^2}^- = \mathcal{L}_{\chi}^- = \frac{1}{2}E_4~, \quad \mathcal{L}^-_{\mathbb{T}(\log\bar{\mathbf{\Phi}})} = 0~.
\label{eq:susy:invariants3}
\end{equation}
We have hence shown that, when considering Einstein-Maxwell solutions in $\mathcal{N}=2$ supergravity, he supersymmetrized Weyl and Euler invariants coincide, while the supersymmetric invariant corresponding to the $\mathbb{T}(\log\bar{\mathbf{\Phi}})$ multiplet becomes trivial.

We first note that all field strength terms have dropped out of the allowed four-derivative Lagrangians (\ref{eq:susy:invariants3}).  This was expected, based on how the analysis of section~\ref{sec:duality} showed that electromagnetic duality prohibits such terms.  However, the duality analysis allowed for the possibility of independent  $W^2$ and $E_4$ terms in the four-derivative Lagrangian, since both terms are purely geometric.

What we have shown in (\ref{eq:susy:invariants3}) is that supersymmetry does not allow for a $W^2$ term in the four-derivative action.  Both the supersymmetrized Euler and supersymmetrized Weyl invariants coincide on-shell with the ordinary Euler invariant.  Supersymmetry is therefore responsible for drastic simplifications to the four-derivative action, even for solutions that do not preserve any supersymmetries of the theory itself.  This generalizes the results of~\cite{Charles:2016wjs}, where it was shown that the supersymmetrized Weyl invariant coincides with the Gauss-Bonnet invariant for Einstein-Maxwell solutions to minimal $\mathcal{N}=2$ supergravity.

In summary, we have shown that the $c$-anomaly must vanish for Einstein-Maxwell solutions embedded in $\mathcal{N}=2$ supergravity: supersymmetry at the level of 
the effective action guarantees that no $W^2$ term can appear. The result applies to each  individual $\mathcal{N}=2$ multiplet by itself and confirms explicit computations in~\cite{Charles:2015eha}. It applies for any Einstein-Maxwell solutions, including those that are not supersymmetric.  As discussed in section~\ref{sec:effQFT:entropy}, the logarithmic corrections to black hole entropy are therefore topological.  In particular, they are independent of continuous parameters such as the black hole mass.  The coefficient of the logarithmic correction remains the same as we deform a supersymmetric black hole off extremality and break supersymmetry by any amount.

\section*{Acknowledgements}
We thank Ratindranath Akhoury, Jan de Boer, Clifford Cheung, Henriette Elvang, Tom Hartman, and Bernard de Wit for useful discussions.  This work was supported in part by the U.S. Department of Energy under grant DE-FG02-95ER40899.

\appendix
 
\section{Notation}
\label{sec:notation}

The setting is a 4D Lorentzian spacetime, with $(-+++)$ signature, where spacetime indices (also known as curved space indices) are denoted by $\mu,\nu,\ldots$ and flat tangent space indices by $a,b,\ldots\,$.  Many of the fields of consideration will also be charged under an $SU(2)$ gauge group, and we will denote the corresponding $SU(2)$ indices of these fields by $i,j,\ldots\,$.  We denote antisymmetrized and symmetrized indices by
\begin{equation}
	[\mu\nu] = \frac{1}{2}(\mu\nu - \nu\mu)~,\quad (\mu\nu) = \frac{1}{2}(\mu\nu + \nu\mu)~,
\end{equation}
with similar expressions for tangent space indices and $SU(2)$ indices.

The spacetime metric is $g_{\mu\nu}$ and the flat space metric is $\eta_{ab}$.  The two are related via the vierbein $e_\mu^{~a}$, allowing conversion between tangent space indices and curved space indices on any Lorentz tensor.  As such, we will be casual about whether we use flat or curved indices.  The only time where the distinction is important is in determining how the supercovariant derivative acts, as it acts non-trivially on the vierbein, and thus the supercovariant derivative acts differently on tensors in flat space differently than tensors in curved space.

We will also make extensive use of the Levi-Civita tensor $\varepsilon_{\mu\nu\rho\sigma}$, a totally anti-symmetric tensor normalized by
\begin{equation}
	\varepsilon_{0123} = \sqrt{-g}~, \quad \varepsilon^{0123} = -\frac{1}{\sqrt{-g}}~.
\end{equation}
In flat Minkowski space, the metric determinant is $\sqrt{-g}=1$, and so this simply becomes the usual Lorentzian Levi-Civita symbol.  The Levi-Civita tensor satisfies the contraction identity
\begin{equation}
	\varepsilon_{\mu_1\ldots\mu_n \nu_1\ldots\nu_p }\varepsilon^{\mu_1\ldots\mu_n\rho_1\ldots\rho_p} = -n!\,p!\, \delta_{\nu_1}^{[\rho_1}\ldots\delta_{\nu_p}^{\rho_p]}~.
\end{equation}
For a $U(1)$ field strength $F_{\mu\nu}$, we will denote the dual field strength by
 \be \label{eq:N2:tF}\tilde{F}_{\mu\nu} \equiv -\frac{i}{2}\varepsilon_{\mu\nu\rho\sigma}F^{\rho\sigma}~.\ee
We can also express the (anti-)self-dual parts of this field strength as
 \be \label{eq:N2:Fpm} F^{\pm}_{\mu\nu} \equiv \frac12\left( F_{\mu\nu} \pm \tilde{F}_{\mu\nu}\right)~.\ee
\section{Off-Shell 4D $\mathcal{N}=2$ Supergravity}
\label{sec:appN2}

In this section, we summarize some of the important technical details of four-dimensional $\mathcal{N}=2$ supergravity in the off-shell formalism.  These details have been studied extensively in previous works~\cite{Mohaupt:2000mj,deWit:1980lyi,deWit:1979dzm,LopesCardoso:2000qm,deWit:1983xhu,deWit:1984rvr}.  We will first discuss the construction of the relevant supersymmetry multiplets and then go into detail discussing the bosonic part of the $\mathcal{N}=2$ conformal supergravity action that couples these multiplets together, complete with higher-derivative interactions.  We then go on to show how, through appropriate gauge-fixing, we can obtain a Poincar\'e supergravity action. We conclude with the consistent bosonic truncation that we make use of in this work.

\subsection{$\mathcal{N}=2$ Superconformal Gravity and the Weyl Multiplet}
\label{sec:appN2:gens}

We first want to construct an $\mathcal{N}=2$ superconformal gauge theory in which all of the generators act as internal symmetries.  To do so, we can take the generators of the $\mathcal{N}=2$ superconformal algebra and introduce a gauge field associated with each generator.  These generators and associated gauge fields are given in table~\ref{tab:supersym}.
	
	\bgroup
	\def\arraystretch{1.5}
	\begin{table}[h]
	\centering
	\begin{tabular}{|c|c|c|}\hline
	{\bf Transformation} & {\bf Generator} & {\bf Gauge Field} \\ \hline
	Translations & $P_a$ & $e_\mu^{~a}$ \\ \hline
	Lorentz & $M_{ab}$ & $\omega_\mu^{ab}$ \\ \hline
	Dilatations & $D$ & $b_\mu$ \\ \hline
	Special conformal & $K_a$ & $f_\mu^{~a}$ \\ \hline
	$SU(2)_R$ & $V^{~j}_{i}$ & $\mathcal{V}_{\mu~j}^{~i}$ \\ \hline
	$U(1)_R$ & $A$ & $A_\mu$ \\ \hline
	$Q$-supersymmetry & $Q^i$ & $\psi_\mu^i$ \\ \hline
	$S$-supersymmetry & $S^i$ & $\phi_\mu^i$ \\ \hline	
	\end{tabular}
	\caption{$\mathcal{N}=2$ superconformal symmetries and their corresponding generators in the $\mathcal{N}=2$ superconformal algebra, as well as the gauge fields associated with each transformation.}
	\label{tab:supersym}
	\end{table}
	\egroup	
	
In principle, we need to define a derivative operator $D_\mu$ that is covariant with respect to the full set of $\mathcal{N}=2$ superconformal symmetries.  Acting with the fully supercovariant derivative on fields can in general yield very lengthy and complicated expressions due to the multitude of gauge fields.  We can define a new, simpler derivative operator $\mathcal{D}_\mu$ that is covariant with respect to Lorentz transformations, dilatations, $R$-symmetry transformations, and whatever other internal gauge transformations the field transforms under.  For example, if $\phi^{\mu_1\ldots\mu_n}$ is a bosonic field with a Weyl weight $w$, a chiral $U(1)_R$ weight $c$, and no $SU(2)_R$ charge, the covariant derivative $\mathcal{D}_\mu$ acts on $\phi^{\mu_1\ldots\mu_n}$ by
\begin{equation}
	\mathcal{D}_{\mu} \phi^{\mu_1\ldots\mu_n} = (\nabla_\mu - w b_\mu - i c A_\mu)\phi^{\mu_1\ldots\mu_n}~,
\label{eq:covderiv_def}
\end{equation}
where $\nabla_\mu$ is the ordinary covariant derivative in curved space with respect to Lorentz transformations.  We will eventually gauge-fix such that we obtain a Poincar\'e supergravity theory in section~\ref{sec:appN2:gauge} and then truncate the theory such that all fermions and $SU(2)_R$-charged fields are set to zero in section~\ref{sec:appN2:trunc}, all of which will make the covariant derivative (\ref{eq:covderiv_def}) more useful than the full supercovariant derivative $D_\mu$.

To now obtain a conformal supergravity theory, the superconformal symmetries must be realized as spacetime symmetries instead of internal ones.  This leads to the (conventional) constraints that make the fields
\begin{equation}
	\omega_\mu^{ab}~, \quad \phi_\mu^i~, \quad f_\mu^{~a}~,
\end{equation}
into composite fields.  In doing so, we are forced to introduce new auxiliary degrees of freedom in the form of an anti-self-dual tensor $T^-_{ab}$, an $SU(2)$ doublet of Majorana spinors $\chi^i$, and a real scalar field $D$\footnote{At this point, we have presented the $R$-symmetry gauge fields as real, physical fields.  However, the $SU(2)_R$ gauge field will eventually be gauge-fixed to zero, and the $U(1)_R$ gauge field does not have a kinetic term at two-derivative order in the action.  We can therefore, from the perspective of the on-shell $\mathcal{N}=2$ supergravity formalism, consider these to be auxiliary fields with no true dynamical degrees of freedom.}.

The remaining independent gauge fields, along with these new auxiliary degrees of freedom, form a superconformal gauge multiplet known as the Weyl multiplet.  The Weyl multiplet, introduced in (\ref{eqn:Weylmultiplet}), can be represented as
\begin{equation}
	\left(\,e_\mu^{~a}~,~\psi_\mu^i~,~b_\mu~,~A_\mu~,~\mathcal{V}_{\mu~j}^{~i}~,~T^-_{\mu\nu}~,~\chi^i~,~D~\right)~,
	\label{eqn:Weylmultiplet_app}
\end{equation}
with 24+24 off-shell bosonic and fermionic degrees of freedom.  

\subsection{Other $\mathcal{N}=2$ Superconformal Multiplets}

We now want to introduce matter in the form of other superconformal multiplets.  In this section, we will detail the field content of the vector, chiral, and non-linear multiplets.

The first multiplet we will consider is the vector multiplet given in (\ref{eq:vectormult}).  It is denoted as
\begin{equation}
	\mathbf{X}^I = \left(\,X^I~,~\Omega^I_i~,~W_\mu^I~,~Y^I_{ij}~\right)~,
\label{eq:vectormult_app}
\end{equation}
with 8+8 off-shell bosonic and fermionic degrees of freedom in the form of a complex scalar $X^I$, an $SU(2)$ doublet of chiral gauginos $\Omega_i^I$, a vector field $W_\mu^I$, and an auxiliary $SU(2)$ triplet of real scalars $Y^I_{ij}$.  These vector multiplets are indexed by $I$.  We need at least one in the theory in order to have enough degrees of freedom to gauge-fix down to Poincar\'e supergravity.  From the perspective of the on-shell formalism, one of the vector multiplets will get combined with the Weyl multiplet to form a gravity multiplet, while the remaining off-shell vector multiplets will become physical vector multiplets.  We therefore let the index $I$ range over
\begin{equation}
	I = 0,\ldots,n_V~,
\end{equation}
where $n_V$ is the number of physical vector multiplets we want to couple to the gravity multiplet.

The next multiplet we will consider is the chiral multiplet, introduced in (\ref{eq:chiralmult}).  The field content of the chiral multiplet is
\begin{equation}
	\hat{\mathbf{A}} = \left(\,\hat{A}~,~\hat{\Psi}_i~,~\hat{B}_{ij}~,~\hat{F}^-_{ab}~,~\hat{\Lambda}_i~,~\hat{C}~\right)~,
\label{eq:chiralmult_app}
\end{equation}
with $16+16$ off-shell degrees of freedom in the form of the complex scalars $\hat{A}$ and $\hat{C}$, $SU(2)$ doublets of left-handed fermions $\hat{\Psi}_i$ and $\hat{\Lambda}_i$, an $SU(2)$ triplet of complex scalars $\hat{B}_{ij}$, and an anti-self-dual tensor $\hat{F}^-_{ab}$ that is antisymmetric in its indices\footnote{We write this anti-self-dual tensor in (\ref{eq:chiralmult}) as $\hat{F}^-_{\mu\nu}$ instead of $\hat{F}^-_{ab}$.  In doing so, we have implicitly converted from tangent space indices to curved space indices via use of the vierbein $e_\mu^{~a}$.}.  The chiral multiplet can in principle be an independent multiplet, but we will eventually consider it to be a composite function of the Weyl and vector multiplet fields in order to introduce higher-derivative interactions into the action.

The last multiplet we will discuss here is the non-linear multiplet, denoted as
\begin{equation}
	\left(\,\Phi^i_{~\alpha}~,~\lambda^i~,~M^{ij}~,~V_a\,\right)~.
\label{eq:nlmult}
\end{equation}
The non-linear multiplet consists of an $SU(2)$ matrix scalar fields $\Phi^i_{~\alpha}$ (where $i$ is the $SU(2)_R$ index and $\alpha = 1,2$ is an additional rigid $SU(2)$ index), a spinor doublet $\lambda^i$, an antisymmetric matrix of complex scalars $M^{ij}$, and a real vector field $V_a$.  The constraint
\begin{equation}
	\mathcal{D}^\mu V_\mu -\frac{1}{2}V^\mu V_\mu - \frac{1}{4}|M_{ij}|^2 + \mathcal{D}^\mu \Phi^i_{~\alpha}\mathcal{D}_\mu \Phi^\alpha_{~i} +\text{(fermions)} = D + \frac{1}{3}R
\label{eq:nlconstraint}
\end{equation}
must be imposed on the non-linear multiplet fields to assure that the multiplet has the correct $8+8$ off-shell degrees of freedom.

\subsection{Prepotential and the Action}
	
	In the previous section, we constructed superconformal multiplets that each transform under some representation of the full $\mathcal{N}=2$ superconformal group.  In particular, we discussed the Weyl multiplet, vector multiplets, chiral multiplets, and non-linear multiplets.  We now want a theory that couples together the Weyl multiplet to $n_V+1$ vector multiplets and a single chiral multiplet.  That is, we would like an action that couples all of these multiplets together such that the $\mathcal{N}=2$ superconformal symmetry is preserved.
	
	One of the ways to accomplish this is to specify the interactions between the Weyl multiplet and the matter fields in the vector and chiral multiplets by introducing a prepotential $F\equiv F(X^I,\hat{A})$, a meromorphic function of the vector multiplet scalars $X^I$ and the chiral multiplet scalar $\hat{A}$.  Derivatives of the prepotential are denoted by
\begin{equation}
	\frac{\partial F}{\partial X^I} = F_I~,\quad \frac{\partial F}{\partial \hat{A}} = F_A~.
\end{equation}
The prepotential is holomorphic and does not depend on the complex conjugate scalars $\bar{X}^I$ and $\bar{\hat{A}}$, and so $F_{\bar{I}} = F_{\bar{A}} = 0$.  The prepotential is also homogeneous of second degree with respect to Weyl-weighted scalings of $X^I$ and $\hat{A}$, so
\begin{equation}
	F(\lambda X^I,\lambda^w \hat{A}) = \lambda^2 F(X^I,\hat{A})~,
\label{eq:prepot2}
\end{equation}
where $w$ is the Weyl weight of the chiral multiplet scalar $\hat{A}$ and $\lambda$ is some arbitrary scaling constant. 

The action is
\begin{equation}
	I = \int d^4x\,\sqrt{-g}\,\mathcal{L}~,
\end{equation}
where $\mathcal{L}$ is the Lagrangian for our off-shell theory that couples the Weyl multiplet, the vector multiplets, and the chiral multiplet via interactions dictated by the prepotential:
\begin{equation}\begin{aligned}
	8\pi\mathcal{L} &= \bigg{[} i \mathcal{D}^\mu F_I \mathcal{D}_\mu \bar{X}^I - i F_I \bar{X}^I\left(\frac{1}{6}R - D \right) -\frac{i}{8}F_{IJ}Y_{ij}^I Y^{Jij} \\
	&\quad + \frac{i}{4}F_{IJ}\left(F_{\mu\nu}^{-I} - \frac{1}{4}\bar{X}^I T_{\mu\nu}^-\right)\left(F^{-\mu\nu J} - \frac{1}{4}\bar{X}^J T^{-\mu\nu}\right) \\
	&\quad - \frac{i}{8} F_I \left(F_{\mu\nu}^{+I} - \frac{1}{4}X^I T_{\mu\nu}^+\right)T^{+\mu\nu} - \frac{i}{32}F T_{\mu\nu}^+ T^{+\mu\nu} \\
	&\quad + \frac{i}{2}F_{AI}\left(F_{\mu\nu}^{-I} - \frac{1}{4}\bar{X}^I T_{\mu\nu}^-\right)\hat{F}^{-\mu\nu} - \frac{i}{4} F_{AI}\hat{B}_{ij}Y^{Iij} \\
	&\quad + \frac{i}{2} F_A \hat{C} - \frac{i}{8}F_{AA}\hat{B}_{ij}\hat{B}_{kl}\varepsilon^{ik}\varepsilon^{jl} + \frac{i}{4}F_{AA} \hat{F}_{\mu\nu}^- \hat{F}^{-\mu\nu} \bigg{]} +\text{h.c.} \\
	&\quad + \text{(fermions)}~.
\label{eq:l_full}
\end{aligned}\end{equation}
We will eventually be interested in purely bosonic backgrounds, so we do not need the details of the fermionic terms.  The covariant derivative $\mathcal{D}_\mu$ defined in (\ref{eq:covderiv_def}) acts on the vector multiplet scalars $X^I$ and the chiral multiplet scalar $\hat{A}$ by
\begin{equation}
	\mathcal{D}_\mu X^I = (\partial_\mu - b_\mu +i A_\mu)X^I~, \quad \mathcal{D}_\mu \hat{A} = (\partial_\mu - w b_\mu + i w A_\mu)\hat{A}~.
\label{eq:covderiv}
\end{equation}

The Lagrangian (\ref{eq:l_full}) has a term linear in the auxiliary $D$ field
\begin{equation}
	8\pi\mathcal{L} = i(F_I \bar{X}^I - \bar{F}_I X^I)\left(D-\frac{1}{6}R\right) + ...~,
\end{equation}
which leads to inconsistent equations of motion.  In order to fix this, we can couple the theory to the non-linear multiplet (\ref{eq:nlmult}) such that all linear terms in $D$ are cancelled.  We add the term
\begin{equation}
	i(F_I \bar{X}^I - \bar{F}_I X^I)\left(\mathcal{D}^\mu V_\mu -\frac{1}{2}V^\mu V_\mu - \frac{1}{4}|M_{ij}|^2 + \mathcal{D}^\mu \Phi^i_{~\alpha}\mathcal{D}_\mu \Phi^\alpha_{~i} - D - \frac{1}{3}R\right)
\end{equation}
to the Lagrangian, modulo some fermionic terms.  The non-linear multiplet constraint (\ref{eq:nlconstraint}) makes this vanish, allowing us to consistently add it to the Lagrangian and cancel out all explict $D$-terms in (\ref{eq:l_full}).  The resulting Lagrangian is
\begin{equation}\begin{aligned}
	8\pi\mathcal{L} &= -\frac{i}{2} (F_I \bar{X}^I - \bar{F}_I X^I)R + \bigg{[} i \mathcal{D}^\mu F_I \mathcal{D}_\mu \bar{X}^I -\frac{i}{8}F_{IJ}Y_{ij}^I Y^{Jij} \\
	&\quad + \frac{i}{4}F_{IJ}\mathcal{F}^{-I}_{\mu\nu}\mathcal{F}^{-\mu\nu J} - \frac{i}{8} F_I \mathcal{F}^{+I}_{\mu\nu}T^{+\mu\nu} - \frac{i}{32}F T_{\mu\nu}^+ T^{+\mu\nu} + \frac{i}{2}F_{AI}\mathcal{F}^{-I}_{\mu\nu}\hat{F}^{-\mu\nu} \\
	&\quad - \frac{i}{4}F_{AI}\hat{B}_{ij}Y^{Iij} + \frac{i}{2} F_A \hat{C} - \frac{i}{8}F_{AA}\hat{B}_{ij}\hat{B}_{kl}\varepsilon^{ik}\varepsilon^{jl} + \frac{i}{4} \hat{F}_{\mu\nu}^- \hat{F}^{-\mu\nu}\bigg{]} +\text{h.c.} \\
	&\quad + i(F_I \bar{X}^I - \bar{F}_I X^I)\left(\mathcal{D}^\mu V_\mu -\frac{1}{2}V^\mu V_\mu - \frac{1}{4}|M_{ij}|^2 + \mathcal{D}^\mu \Phi^i_{~\alpha}\mathcal{D}_\mu \Phi^\alpha_{~i} \right) \\
	&\quad + \text{(fermions)}~,
\label{eq:ltot}
\end{aligned}\end{equation}
where we have defined the supercovariant field strengths
\begin{equation}\begin{aligned}
	\mathcal{F}^{+I}_{\mu\nu} &= F^{+I}_{\mu\nu} - \frac{1}{4} X^I T^+_{\mu\nu}~, \\
	\mathcal{F}^{-I}_{\mu\nu} &= F^{-I}_{\mu\nu} - \frac{1}{4} \bar{X}^I T^-_{\mu\nu}~.
\label{eq:superf}
\end{aligned}\end{equation}

\subsection{Introducing Higher-Derivative Terms}
\label{sec:appN2:higher}

We are interested in studying higher-derivative interactions in $\mathcal{N}=2$ supergravity.  As discussed in section~\ref{sec:form:fourderiv}, we can accomplish this by identifying the chiral multiplet (\ref{eq:chiralmult_app}) as a composite multiplet of other fields.  In this section, we will discuss the two known chiral multiplets that introduce four-derivative terms into the action.
	
\subsubsection{$\mathbf{W}^2$ Multiplet}
	
	The fields in the Weyl multiplet can be also be fit into a chiral multiplet, denoted as
\begin{equation}
	\mathbf{W}_{ab} = \left(A_{ab}\,,\,\Psi_{abi}\,,\,B_{abij}\,,\,(F^-_{ab})_{cd}\,,\,\Lambda_{abi}\,,\,C_{ab}\right)~,
\end{equation}
of which the bosonic components are
\begin{equation}\begin{aligned}
	A_{ab}|_{\mathbf{W}_{ab}} &= T^-_{ab}~, \\
	B_{abij}|_{\mathbf{W}_{ab}}	&= -8 \varepsilon_{k(i}\mathcal{V}_{ab~~j)}^{-~k}~, \\
	(F^-_{ab})^{cd}|_{\mathbf{W}_{ab}} &= -8W^{-~cd}_{ab} - 4 \left(\delta_{[a}^{~c}\delta_{b]}^{~d} + \frac{i}{2}\varepsilon_{ab}^{~~cd}\right)D + 16 i A^{-[c}_{[a}\delta_{b]}^{~d]}~, \\
	C_{ab}|_{\mathbf{W}_{ab}} &= 4 D_{[a}D^c T^+_{b]c} + 4 D^c D_{[a}T^+_{b]c} + 2 \square_c T^+_{ab} ~.
\end{aligned}\end{equation}
where we have defined
\begin{equation}
	W^{-~cd}_{ab} = \frac{1}{2}\left(W_{ab}^{~~cd} + i^*W_{ab}^{~~cd}\right)~,\quad {}^*W_{ab}^{~~cd} = \frac{1}{2}\varepsilon_{ab}^{~~ef}W_{ef}^{~~cd}~.
\end{equation}
We can then obtain the chiral multiplet $\mathbf{W}^2$ by squaring $\mathbf{W}_{ab}$, i.e.
\begin{equation}
	\mathbf{W}^2 = \mathbf{W}_{ab}\mathbf{W}^{ab}~,
\end{equation}
where chiral multiplets are multiplied using superconformal calculus rules discussed in~\cite{deWit:2010za}.  The bosonic components of $\mathbf{W}^2$ are
\begin{equation}\begin{aligned}
	A|_{\mathbf{W}^2} &= T^-_{ab} T^{-ab}~, \\
	B_{ij}|_{\mathbf{W}^2} &= -16 \varepsilon_{k(i}\mathcal{V}_{ab~j)}^{~~k}T^{-ab}~, \\
	F^-_{ab}|_{\mathbf{W}^2} &= -16\left( W_{abcd}T^{-cd} +D T^-_{ab} + 2 i A_{c[a}T^{-c}_{b]}\right)~, \\
	C|_{\mathbf{W}^2} &= 32\left(\vphantom{\frac{1}{1}} W_{abcd}W^{abcd} + i^* W_{abcd}W^{abcd} + 6D^2 - 2 A_{ab}A^{ab} \right. \\
	&\quad\quad + 2 A_{ab}\tilde{A}^{ab} - \frac{1}{2}T^{-ab} \mathcal{D}_a \mathcal{D}^c T^+_{cb} + \frac{1}{4}R^a_{~b}T^-_{ac}T^{+bc} \\
	&\left.\quad\quad + \frac{1}{256}T^-_{ab}T^{-ab}T^+_{cd}T^{+cd} + \frac{1}{2}\mathcal{V}_{ab~j}^{~~i}\mathcal{V}^{ab j}_{~~~~i} - \frac{1}{2}\mathcal{V}_{ab~j}^{~~i}\tilde{\mathcal{V}}^{ab j}_{~~~~i} \right)~.
\label{eq:chiral_w2}
\end{aligned}\end{equation}
The scalar $C|_{\mathbf{W}^2}$ in the $\mathbf{W}^2$ multiplet has $(\text{Weyl})^2$-type terms in it.  This introduces four-derivative terms into the Lagrangian (\ref{eq:ltot}), making the $\mathbf{W}^2$ chiral multiplet one way to introduce higher-derivative terms into $\mathcal{N}=2$ supergravity.	
	
\subsubsection{$\mathbb{T}(\log\bar{\mathbf{\Phi}})$ Multiplet}
	
	Let $\mathbf{\Phi}$ be an arbitrary chiral multiplet, denoted by
\begin{equation}
	\mathbf{\Phi} = \left(A\,,\,\Psi_i\,,\,B_{ij}\,,\,F^-_{ab}\,,\,\Lambda_i\,,\,C\,\right)~,
\end{equation}
The Hermitian conjugate of $\mathbf{\Phi}$ is the anti-chiral multiplet $\bar{\mathbf{\Phi}}$, denoted by
\begin{equation}
	\bar{\mathbf{\Phi}} = \left(\bar{A}\,,\,\Psi^i\,,\,B^{ij}\,,\,F^+_{ab}\,,\,\Lambda^i\,,\,\bar{C}\,\right)~.
\end{equation}
From the chiral multiplet $\mathbf{\Phi}$, we can also construct the chiral multiplet $\log\mathbf{\Phi}$.  Ignoring all fermions, the bosonic components of $\log\mathbf{\Phi}$ are related those of $\mathbf{\Phi}$ by
\begin{equation}\begin{aligned}
	A|_{\log \mathbf{\Phi}} &= \log A|_{\mathbf{\Phi}}~, \\
	B_{ij}|_{\log \mathbf{\Phi}} &= \frac{B_{ij}}{A}\bigg{|}_{\mathbf{\Phi}}~, \\
	F^-_{ab}|_{\log \mathbf{\Phi}} &= \frac{F^-_{ab}}{A}\bigg{|}_{\mathbf{\Phi}}~, \\
	C|_{\log \mathbf{\Phi}} &= \left(\frac{C}{A} + \frac{1}{4 A^2}\left( \varepsilon^{ik}\varepsilon^{jl}B_{ij}B_{kl}-2F^-_{ab}F^{-ab}\right)\right) \bigg{|}_{\mathbf{\Phi}}~.
\label{eq:logchiral}
\end{aligned}\end{equation}
We can also take the Hermitian conjugate of this multiplet to obtain the anti-chiral multiplet $\log \bar{\mathbf{\Phi}}$.  We can then construct the chiral kinetic multiplet $\mathbb{T}(\log \bar{\mathbf{\Phi}})$ whose bosonic components are related to the components of $\log \bar{\mathbf{\Phi}}$ by
\begin{equation}\begin{aligned}
	A|_{\mathbb{T}(\log \bar{\mathbf{\Phi}})} &= \bar{C}|_{\log\bar{\mathbf{\Phi}}}~, \\
	B_{ij}|_{\mathbb{T}(\log \bar{\mathbf{\Phi}})} &= \left(-2 \varepsilon_{ik}\varepsilon_{jl}(\square_c + 3 D)B^{kl} - 2\varepsilon_{jk} \mathcal{V}^{ab~k}_{~~~~~i} F^+_{ab}\right)|_{\log\bar{\mathbf{\Phi}}}~, \\
	F^-_{ab}|_{\mathbb{T}(\log \bar{\mathbf{\Phi}})} &= \bigg{(}T^-_{ab} \square_c \bar{A} - \varepsilon_{ij}\mathcal{V}_{ab~k}^{-~i}B^{jk} + \frac{1}{16}T^-_{ab}T^+_{cd}F^{+cd} \\
	&\quad -\Pi_{ab}^{-~cd} \left(4 D_c D^e F^+_{ed} + (D_c T^-_{de})D^e \bar{A} + (D^e T^-_{ed}) D_c \bar{A}  - w D_c D^e T^-_{ed}\right)\bigg{)}\bigg{|}_{\log\bar{\mathbf{\Phi}}}~,\\
	C|_{\mathbb{T}(\log \bar{\mathbf{\Phi}})} &= \left(4(\square_c + 3D)\square_c \bar{A} + 6 (D_a D) D^a \bar{A} - 16 D^a\left(R(D)^+_{ab} D^b \bar{A}\right)\frac{}{}\right. \\
	&\quad -\frac{1}{2}D^a(T^+_{ab}T^{-cb}D_c \bar{A}) - \frac{1}{4}D^a(T^+_{ab}T^{-cb})D_c \bar{A} + \frac{1}{16}T^+_{ab}T^{+ab}\bar{C} \\
	&\quad+ \frac{1}{2}\square_c (T^+_{bc}F^{+bc}) + 2 D_a\left((D^b T^+_{bc})F^{+ac} + T^{+ac} D^b F^+_{bc}\right)-w \mathcal{V}^{+~i}_{ab~j}\mathcal{V}^{+ab~j}_{~~~~i} \\
	&\quad\left. - 8 w R(D)^+_{ab}R(D)^{+ab} -\frac{w}{2} D^a T^+_{ab}D_c T^{-cb} -\frac{w}{2} D^a(T^+_{ab}D_c T^{-cb})\right)\bigg{|}_{\log\bar{\mathbf{\Phi}}}~,
\label{eq:kineticlog}
\end{aligned}\end{equation}
where $w$ is the Weyl weight of the $\mathbf{\Phi}$ multiplet, $\Pi^{-~cd}_{ab}$ is the anti-self-dual projection operator
\begin{equation}
	\Pi^{-~cd}_{ab} = \delta_a^{~[c}\delta_b^{~d]} + \frac{i}{2}\varepsilon_{ab}^{~~cd}~,
\end{equation} and $R(D)^+_{ab}$ is the self-dual part of the connection $R(D)_{ab}$ defined by
\begin{equation}
	R(D)_{\mu\nu} = 2 \partial_{[\mu}b_{\nu]} + \frac{i}{2}\tilde{A}_{\mu\nu}~.
\label{eq:r(d)}
\end{equation}
Note that the derivative operator $D_\mu$ appearing in (\ref{eq:kineticlog}) is the fully superconformally covariant derivative discussed in section~\ref{sec:appN2:gens}, and the operator $\square_c \equiv D_\mu D^\mu$ is the superconformal d'Alembertian.  These can be expressed in terms of the covariant derivative $\mathcal{D}_\mu$ and its square $\mathcal{D}^2$~\cite{deWit:2010za}.  For our purposes, though, we will only need the particular linear combination of these fields appearing in (\ref{eq:susy:tlogmultiplet}), which simplifies in such a way that no explicit occurences of the superconformal derivative appear.
	
\subsubsection{Higher-Derivative Action}

	The Poincar\'e supergravity Lagrangian (\ref{eq:ltot}) couples an arbitrary chiral multiplet $\hat{\mathbf{A}}$ to the Weyl and vector multiplets.  By identifying this chiral multiplet with a linear combination of $\mathbf{W}^2$ and $\mathbb{T}(\log\bar{\Phi})$, both of which contain four-derivative terms, the Lagrangian will contain (at least) four-derivative terms in it. That is, we will set
\begin{equation}
	\hat{\mathbf{A}} = a_1 \mathbf{W}^2 + a_2 \mathbb{T}(\log\bar{\mathbf{\Phi}})~,
\label{eq:ahatsum}
\end{equation}
for some constants $a_1,a_2$.  This sets the bosonic components of $\hat{\mathbf{A}}$ to be
\begin{equation}\begin{alignedat}{2}
	\hat{A} &= a_1 A|_{\mathbf{W}^2}&				&+ a_2 A|_{\mathbb{T}(\log\bar{\mathbf{\Phi}})} \\
	\hat{B}_{ij} &= a_1 B_{ij}|_{\mathbf{W}^2}&		&+ a_2 B_{ij}|_{\mathbb{T}(\log\bar{\mathbf{\Phi}})} \\
	\hat{F}^-_{ab} &= a_1 F^-_{ab}|_{\mathbf{W}^2}&	&+ a_2 F^-_{ab}|_{\mathbb{T}(\log\bar{\mathbf{\Phi}})} \\
	\hat{C} &= a_1 C|_{\mathbf{W}^2}&				&+ a_2 C|_{\mathbb{T}(\log\bar{\mathbf{\Phi}})}
\label{eq:aident_comp}
\end{alignedat}\end{equation}
Under this identification, the prepotential $F(X^I,\hat{A})$ is a function of the scalars $X^I$ and $\hat{A} = a_1 A|_{\mathbf{W}^2} + a_2 A|_{\mathbb{T}(\log\bar{\mathbf{\Phi}})}$.  Both $A|_{\mathbf{W}^2}$ and $A|_{\mathbb{T}(\log\bar{\mathbf{\Phi}})}$ have Weyl weight $w = 2$, and so the homogeneity relation of the prepotential (\ref{eq:prepot2}) tells us that
\begin{equation}
	F_I X^I + 2 F_A \hat{A} = 2 F~.
\end{equation}

The Lagrangian (\ref{eq:ltot}), subject to the identification (\ref{eq:aident_comp}), can now contain higher-derivative terms, with the derivative order depending on the form of the prepotential.  We will represent the prepotential perturbatively as
\begin{equation}\begin{aligned}
	F(X^I, \hat{A}) &= \sum_{n=0}^\infty F^{(n)}(X^I) \hat{A}^n \\
	&= F^{(0)}(X^I) + F^{(1)}(X^I) \hat{A} + \ldots~,
\label{eq:pertprepot}
\end{aligned}\end{equation}
for some functions $F^{(n)}(X^I)$.  The zeroth-order function $F^{(0)}(X^I)$ dictates the two-derivative terms in the Lagrangian, the first order function $F^{(1)}(X^I)$ dictates the four-derivative terms in the Lagrangian, and so on. As discussed in section~\ref{sec:form:fourderiv}, we truncate the prepotential to finite order:
\begin{equation}
	F(X^I, \hat{A}) = F^{(0)}(X^I) + F^{(1)}(X^I) \hat{A}~,
\label{eq:prepot_twoterms}
\end{equation}
in order to have only two- and four-derivative interactions.  The bosonic two-derivative part of the Lagrangian is
\begin{equation}\begin{aligned}
	8\pi\mathcal{L}^{(2)} &= -\frac{i}{2} (F^{(0)}_I \bar{X}^I - \bar{F}^{(0)}_I X^I)R + \bigg{[} i \mathcal{D}^\mu F^{(0)}_I \mathcal{D}_\mu \bar{X}^I -\frac{i}{8}F^{(0)}_{IJ}Y_{ij}^I Y^{Jij} \\
	&\quad + \frac{i}{4}F^{(0)}_{IJ}\mathcal{F}^{-I}_{\mu\nu}\mathcal{F}^{-\mu\nu J} - \frac{i}{8} F^{(0)}_I \mathcal{F}^{+I}_{\mu\nu}T^{+\mu\nu} - \frac{i}{32}F^{(0)} T_{\mu\nu}^+ T^{+\mu\nu}\bigg{]} +\text{h.c.} \\
	&\quad + i(F^{(0)}_I \bar{X}^I - \bar{F}^{(0)}_I X^I)\left(\mathcal{D}^\mu V_\mu -\frac{1}{2}V^\mu V_\mu - \frac{1}{4}|M_{ij}|^2 + \mathcal{D}^\mu \Phi^i_{~\alpha}\mathcal{D}_\mu \Phi^\alpha_{~i} \right)~,
\label{eq:l_app_2deriv}
\end{aligned}\end{equation}
while the bosonic four-derivative part is
\begin{equation}\begin{aligned}
	8\pi\mathcal{L}^{(4)} &= -\frac{i}{2} (F^{(1)}_I \bar{X}^I \hat{A} - \bar{F}^{(1)}_I  X^I \bar{\hat{A}})R + \bigg{[} i \mathcal{D}^\mu (F^{(1)}_I \hat{A}) \mathcal{D}_\mu \bar{X}^I -\frac{i}{8}F^{(1)}_{IJ} Y_{ij}^I Y^{Jij} \hat{A} \\
	&\quad + \frac{i}{4}F^{(1)}_{IJ}\mathcal{F}^{-I}_{\mu\nu}\mathcal{F}^{-\mu\nu J}\hat{A} - \frac{i}{8} F^{(1)}_I \mathcal{F}^{+I}_{\mu\nu}T^{+\mu\nu}\hat{A} - \frac{i}{32}F^{(1)} T_{\mu\nu}^+ T^{+\mu\nu}\hat{A} \\
	&\quad + \frac{i}{2}F^{(1)}_{I}\mathcal{F}^{-I}_{\mu\nu}\hat{F}^{-\mu\nu} + \frac{i}{2} F^{(1)} \hat{C} - \frac{i}{4}F^{(1)}_{I}\hat{B}_{ij}Y^{Iij}\bigg{]} +\text{h.c.} \\
	&\quad + i(F^{(1)}_I \bar{X}^I \hat{A} - \bar{F}^{(1)}_I X^I \bar{\hat{A}})\left(\mathcal{D}^\mu V_\mu -\frac{1}{2}V^\mu V_\mu - \frac{1}{4}|M_{ij}|^2 + \mathcal{D}^\mu \Phi^i_{~\alpha}\mathcal{D}_\mu \Phi^\alpha_{~i} \right)~,
\label{eq:l_app_4deriv}
\end{aligned}\end{equation}
subject to the identifications (\ref{eq:aident_comp}) for the chiral multiplet fields.

\subsection{Gauge-Fixing Down to Poincar\'e}
\label{sec:appN2:gauge}
	
	The Lagrangian (\ref{eq:ltot}) has an $\mathcal{N}=2$ superconformal symmetry that acts as an internal symmetry.  To obtain an $\mathcal{N}=2$ Poincar\'{e} supergravity theory, we must gauge-fix the extra symmetries of the superconformal theory, including special conformal transformations, dilatations, and a local chiral $SU(2)_R\, \times \, U(1)_R$ symmetry.  We gauge-fix the special conformal symmetry by choosing the $K$-gauge
\begin{equation}
	b_\mu = 0~.
\label{eq:gauge_k}
\end{equation}
To gauge-fix the dilatational symmetry, we choose the $D$-gauge that sets the K\"{a}hler potential to be constant:
\begin{equation}
	e^{-\mathcal{K}} \equiv i(F_I \bar{X}^I - \bar{F}_I X^I) = \frac{8\pi}{\kappa^2}~,
\label{eq:gauge_d}
\end{equation}
with the value of the constant chosen to reproduce the standard normalization of the Einstein-Hilbert term in the action.  The $SU(2)_R$ symmetry can be gauge-fixed by imposing the $V$-gauge
\begin{equation}
	\Phi^i_{~\alpha} = \delta^i_{~\alpha}~,
\label{eq:gauge_v}
\end{equation}
on the non-linear multiplet, while the $U(1)_R$ symmetry can be gauge-fixed via the $A$-gauge condition
\begin{equation}
	X^0 = \bar{X}^0~.
\label{eq:gauge_a}
\end{equation}
The $D$-gauge (\ref{eq:gauge_d}) and $A$-gauge (\ref{eq:gauge_a}) remove two degrees degree of freedom from the vector multiplet scalars, and thus the Poincar\'{e} supergravity theory has only $n_V$ independent scalars, as expected.  

The gauge choices made here are by no means unique.  And, since physical observables should not depend on the choice of gauge, different sets of gauge-fixing conditions can be useful for different types of problems. The gauge choices presented here are typical and useful in a broad class of applications.
	
\subsection{Consistent Truncation}

\label{sec:appN2:trunc}

We now have a theory of Poincar\'e $\mathcal{N}=2$ supergravity with higher-derivative interactions, introduced by gauging the superconformal symmetries and making particular choices for the chiral multiplet coupled to our theory.  We are in principle at the point where we can solve the full set of equations of motion to our theory and investigate particular solutions.  However, the action presented thus far is fairly complicated and includes implicit dependence on a great number of fields, some physical and some auxiliary.  This makes finding solutions difficult.  We will therefore look at how to consistently truncate our theory down to a more manageable set of fields and interactions.  We will do this by eliminating auxiliary fields from our theory wherever possible.

We are primarily interested in purely bosonic backgrounds, and so we will turn off all fermions.  These backgrounds will still capture the most salient features of our theory, including the structure of black hole entropy corrections.  The fields $Y_{ij}^I$ and $\mathcal{V}_{\mu~j}^{~i}$ and their derivatives couple either to fermionic terms, or appear at least quadratically with one another.  This is true even when higher-derivative terms are present.  It is therefore consistent to set them both to zero
\begin{equation}
	Y_{ij}^I = 0~, \quad \mathcal{V}_{\mu~j}^{~i} = 0~,
\end{equation}
at the level of the action.  Note that this sets the $SU(2)_R$-charged chiral multiplet field $\hat{B}_{ij} = 0$, and so we can ignore all such terms in the action.

Next, we want to eliminate the non-linear multiplet fields $V_\mu$, $M_{ij}$, and $\Phi^i_{~\alpha}$ from our theory.  The scalar fields $\Phi^i_{~\alpha}$ are easy to eliminate: the $V$-gauge condition (\ref{eq:gauge_v}), combined with setting the $SU(2)_R$ gauge field to zero, makes the derivative $\mathcal{D}_\mu \Phi^i_{~\alpha}$ vanish.  The remaining non-linear multiplet fields can be eliminated by noticing that they interact with the other matter fields only through the K\"{a}hler potential $e^{-\mathcal{K}} = i(F_I \bar{X}^I - \bar{F}_I X^I)$, which is set to a constant via the $D$-gauge condition (\ref{eq:gauge_d}).  The non-linear multiplet fields effectively decouple from the rest of our theory, and so we can study their equations of motion independently from the others.  We find that we can choose
\begin{equation}
	V_\mu = 0~, \quad M_{ij} = 0~,
\end{equation}
at the level of the action.  Now that we have eliminated all of the non-linear multiplet fields from the theory, the non-linear multiplet constraint (\ref{eq:nlconstraint}) forces the background value of $D$ to satisfy
\begin{equation}
	D = -\frac{1}{3}R~.
\end{equation}

The only remaining unconstrained auxiliary fields in our theory are the anti-self-dual tensor $T^-_{\mu\nu}$ and the $U(1)_R$ gauge field $A_\mu$.  In principle, we should find their respective equations of motion, solve for these auxiliary fields in terms of physical ones, and then replace them with their on-shell values at the level of the action.  However, this procedure only works when the fields are pure Lagrange multiplier fields with no kinetic terms.  This is spoiled by the higher-derivative interactions introduced in section~\ref{sec:appN2:higher}, which include terms like $T^{-\mu\nu}\mathcal{D}_\mu \mathcal{D}^\rho T^+_{\rho\nu}$ and $A^-_{\mu\nu}A^{-\mu\nu}$.  The equations of motion for these fields are therefore no longer algebraic, and so these auxiliary fields cannot be eliminated in closed form.  However, we take the view that the action is an effective action valid at energy scales well below the UV scale.  We will therefore treat the higher-derivative terms as perturbative corrections to the two-derivative theory, and thus we will still always be able to eliminate all auxiliary fields.

The result of the preceeding discussion, when combined with section~\ref{sec:appN2:gauge}, is that we can eliminate almost all of the fields from our theory.  This is done by imposing the Poincar\'e gauge-fixing conditions
\begin{equation}
	b_\mu = 0~, \quad \Phi^i_{~\alpha} = \delta^i_{~\alpha}~, \quad i(F_I \bar{X}^I - \bar{F}_I X^I) = \frac{8\pi}{\kappa^2}~,
\end{equation}
and then consistently setting 
\begin{equation}
	Y_{ij}^I = \mathcal{V}_{\mu~j}^{~i} = V_\mu = M_{ij} = \text{fermions} = 0~,
\end{equation}
at the level of the action.  The Lagrangian (\ref{eq:ltot}) therefore becomes
\begin{equation}\begin{aligned}
	\mathcal{L} &= -\frac{1}{2\kappa^2}R + \frac{1}{8\pi}\bigg{[} i \mathcal{D}^\mu F_I \mathcal{D}_\mu \bar{X}^I  + \frac{i}{4}F_{IJ}\mathcal{F}^{-I}_{\mu\nu}\mathcal{F}^{-\mu\nu J} - \frac{i}{8} F_I \mathcal{F}^{+I}_{\mu\nu}T^{+\mu\nu}\\
	&\quad  - \frac{i}{32}F T_{\mu\nu}^+ T^{+\mu\nu} + \frac{i}{2}F_{AI}\mathcal{F}^{-I}_{\mu\nu}\hat{F}^{-\mu\nu} + \frac{i}{2} F_A \hat{C} + \frac{i}{4}F_{AA}\hat{F}_{\mu\nu}^- \hat{F}^{-\mu\nu}\bigg{]} +\text{h.c.}~.
\label{eq:l_trunc_app}
\end{aligned}\end{equation}
Additionally, any solution to the equations of motion of this Lagrangian must satisfy the constraints
\begin{equation}
	X^0 = \bar{X}^0~, \quad D = -\frac{1}{3}R~.
\end{equation}
We can also look at the two-derivative and four-derivative parts of the Lagrangian (\ref{eq:l_trunc_app}) by using the two-term prepotential (\ref{eq:prepot_twoterms}).  These are given, respectively, by
\begin{align}
	\mathcal{L}^{(2)} &= -\frac{1}{2\kappa^2}R + \frac{1}{8\pi}\bigg{[} i \mathcal{D}^\mu F^{(0)}_I \mathcal{D}_\mu \bar{X}^I + \frac{i}{4}F^{(0)}_{IJ}\mathcal{F}^{-I}_{\mu\nu}\mathcal{F}^{-\mu\nu J} - \frac{i}{8} F^{(0)}_I \mathcal{F}^{+I}_{\mu\nu}T^{+\mu\nu} \nonumber \\
	&\quad - \frac{i}{32}F^{(0)} T_{\mu\nu}^+ T^{+\mu\nu}\bigg{]} +\text{h.c.}~, \\
	\mathcal{L}^{(4)} &= \frac{1}{8\pi}\bigg{[} i \mathcal{D}^\mu (F^{(1)}_I \hat{A}) \mathcal{D}_\mu \bar{X}^I + \frac{i}{4}F^{(1)}_{IJ}\mathcal{F}^{-I}_{\mu\nu}\mathcal{F}^{-\mu\nu J}\hat{A} - \frac{i}{8} F^{(1)}_I \mathcal{F}^{+I}_{\mu\nu}T^{+\mu\nu}\hat{A} \nonumber \\
	&\quad - \frac{i}{32}F^{(1)} T_{\mu\nu}^+ T^{+\mu\nu}\hat{A} + \frac{i}{2}F^{(1)}_{I}\mathcal{F}^{-I}_{\mu\nu}\hat{F}^{-\mu\nu} + \frac{i}{2} F^{(1)} \hat{C} \bigg{]} +\text{h.c.}~,
\end{align}
which are precisely the Lagrangians presented in (\ref{eq:l_2deriv}) and (\ref{eq:l_4deriv}).

 \section{Duality Calculations}
 In this appendix, we give more details on certain calculations relevant to section \ref{sec:duality}. First, we will give more details of the transformation properties of the four-derivative Lagrangian, as discussed in section \ref{sec:duality:invarianceL4}. Then, we give more details on the calculations in section \ref{sec:duality:fourder}, where the allowed four-derivative duality-invariant terms are studied.

 \subsection{More On Duality Transformation of Four-Derivative Actions}\label{sec:appendix:dualityL4}\label{sec:appendix:transfL4}
We will give more details here for the derivation leading to (\ref{eq:duality:transfL4}) in \ref{sec:appendix:transfL4}. We will also discuss higher-derivative correction to Einstein-Maxwell theory in \ref{sec:appendix:EMexample}, explaining along the way how our results are consistent with earlier work regarding the duality-invariant deformations of Maxwell theory with higher order terms, such as \cite{Bossard:2011ij,Carrasco:2011jv,Broedel:2012gf}.
 
 The derivation leading to (\ref{eq:duality:transfL4}) is a slight generalization of the derivation in appendix B of \cite{Gaillard:1981rj}. Using the notation of \cite{Gaillard:1981rj}, let us consider the general situation with gauge fields (suppressing Lorentz indices) $F^I$, scalars $\chi^i$, and duality transformation functions of the scalars $\delta\chi^i = \xi^i(\chi)$. The duality transformation of a Lagrangian is given by \cite{Gaillard:1981rj}:\footnote{This can easily be generalized to allow for higher derivative terms involving the scalars and gauge fields.}
 \be \label{eq:app:deltaLgen} \delta \mathcal{L} = \left( \xi^i\frac{\partial}{\partial\chi^i} + \partial_{\mu} \xi^i\frac{\partial}{\partial(\partial_{\mu}\chi^i)} + (F^K A^JK + G^K B^{JK})\frac{\partial}{\partial F^J}\right)\mathcal{L}~.\ee
 If in addition the Lagrangian $\mathcal{L}$ gives rise to a duality-invariant theory, then the above must reduce to \cite{Gaillard:1981rj}
 \be \label{eq:app:deltaLdi} \delta \mathcal{L} = \frac14\left( F^J C^{JK}i\tilde{F}^K + G^J B^{JK} i\tilde{G}^K\right).\ee
 Now, assuming that $\mathcal{L}$ depends on a duality-invariant parameter $\lambda$, then we can take the derivative of both sides of (\ref{eq:app:deltaLgen}) with respect to $\lambda$; after some rewriting we get
 \be \delta\left( \frac{\partial\mathcal{L}}{\partial \lambda}\right) = \frac{\partial}{\partial\lambda}\left( \delta \mathcal{L} - \frac14F^J C^{JK}i\tilde{F}^K -\frac14 G^J B^{JK} i\tilde{G}^K\right) - \frac{\partial \xi^i}{\partial\lambda}\frac{\partial\mathcal{L}}{\partial \chi^i} - \partial_{\mu}\left(\frac{\partial\xi^i}{\partial\lambda}\right)\frac{\partial\mathcal{L}}{\partial(\partial_{\mu}\chi^i)}~.\ee
 If duality-invariance is preserved, we can use (\ref{eq:app:deltaLdi}) and conclude
 \be \label{eq:deltalambdaL} \delta\left( \frac{\partial\mathcal{L}}{\partial \lambda}\right) = - \frac{\partial \xi^i}{\partial\lambda}\frac{\partial\mathcal{L}}{\partial \chi^i} - \partial_{\mu}\left(\frac{\partial\xi^i}{\partial\lambda}\right)\frac{\partial\mathcal{L}}{\partial(\partial_{\mu}\chi^i)}~.\ee
 This equation generalizes (B.3) in \cite{Gaillard:1981rj} to the case where the $\xi^i$ are allowed to depend on $\lambda$.
 
 Now, to use (\ref{eq:deltalambdaL}) in the context of four-derivative corrections, assume we have a Lagrangian of the form
 \be \mathcal{L}(\lambda) = \mathcal{L}^{(2)} + \lambda \mathcal{L}^{(4)} + {\cal O}(\lambda^2)~,\ee
 so that $\partial_\lambda(\mathcal{L}) = \mathcal{L}^{(4)}+{\cal O}(\lambda)$. All functions of the fields should be viewed as having a perturbative series in $\lambda$; e.g. the functions $\xi^i$ can be written as
 \be \xi^i = \xi^{(2)i}+\lambda \xi^{(4)i}~.\ee
 Now, it can easily be seen that (\ref{eq:deltalambdaL}) can be rewritten as
 \be \delta \mathcal{L}^{(4)} = - \xi^{(4)i}\frac{\partial\mathcal{L}^{(2}}{\partial \chi^i} - \partial_{\mu}\left(\xi^{(4)i}\right)\frac{\partial\mathcal{L}^{(2)}}{\partial(\partial_{\mu}\chi^i)}~,\ee
 to leading order in $\lambda$. In other words, the duality transformation properties of the subleading piece $\mathcal{L}^{(4)}$ are completely determined (to leading order in $\lambda$) by the leading piece $\mathcal{L}^{(2)}$ and the subleading piece of the duality transformation functions of the scalars $\xi^{(4)i}(\chi)$. Finally, to arrive at (\ref{eq:duality:transfL4}) for the four-derivative corrections of $\mathcal{N}=2$ supergravity, we note that due to the expansion of the prepotential (\ref{eq:Fexpansion}), the functions $F_I$ do indeed have an expansion in $\lambda$, leading to (\ref{eq:duality:transfL4}).

 \subsubsection{Example: Einstein-Maxwell}\label{sec:appendix:EMexample}

As mentioned in section \ref{sec:duality:EinsteinMaxwell}, Maxwell theory with the  Lagrangian (ignoring any coupling to gravity here)
\be \label{eq:app:Maxwell} \mathcal{L} = \frac14 F^2~,\ee
has the duality symmetry given by $SO(2,\mathbb{R})$ rotations of the vector $(F, G)$ with $G$ defined by (\ref{eq:duality:GEM}). If we wish to deform the Lagrangian by adding higher-order (in $F$) terms, the duality vector $(F,G)$ receives corrections due to the definition of $G$. Since the form of the (altered) duality transformations themselves depend on the higher-order terms added to the Lagrangian, it is in principle highly non-trivial to determine what can be added to the Lagrangian while keeping duality invariance of the theory. In fact, it can be proven that if we add any ${\cal O}(F^4)$ terms to the Lagrangian, to ensure duality invariance we would also need to add an infinite amount of higher order terms $F^{2n}$ for all $n>2$ \cite{Bossard:2011ij,Carrasco:2011jv,Broedel:2012gf}. One possible way of doing so is Born-Infeld theory
\be \label{eq:app:BI} \mathcal{L}_{BI} = \frac{1}{g^2} \left(\sqrt{1+2g^2\frac{F^2}{4} + g^4 \frac{F\tilde{F}}{4}^2}-1\right) =  \frac14 F^2 - \frac{1}{8}g^2\left[ F^4 - \frac14(F^2)^2\right] + \cdots,\ee
for which the $g\rightarrow0$ limit clearly gives back the Maxwell action (\ref{eq:app:Maxwell}). There are other non-equivalent ways to deform the Maxwell Lagrangian in a way consistent with duality symmetry \cite{Carrasco:2011jv}.

As we have mentioned many times in this paper, the point of view we are adapting for the ${\cal O}(F^4)$ terms in section \ref{sec:duality} (and in particular section \ref{sec:duality:invarianceL4}) is that these terms are a \emph{perturbative} correction to the two-derivative Lagrangian
\be \mathcal{L} =  \mathcal{L}^{(2)} + g^2 \mathcal{L}^{(4)} + {\cal O}(g^4)~,\ee
so that all relevant quantities must also be expanded consistently in orders of $g$. Thus, to demand duality invariance of our theory at four-derivative order is demanding duality invariance up to order ${\cal O}(g^2)$ only, and not fully non-linear in $g$. For example, keeping only the ${\cal O}(g^2)$ terms in the Born-Infeld action (\ref{eq:app:BI}), we see the unique four-derivative duality-invariant term $\calI_{\mu\ \nu\rho}^{\ \rho} \calI^{\mu\sigma\nu}_{\ \ \ \sigma}$ (see (\ref{eq:duality:MaxwellRic2})) appearing at four-derivative order. This is consistent with our discussion in section \ref{sec:duality:invarianceL4} and above in appendix \ref{sec:appendix:transfL4}, which implies that in a theory without scalars, the four derivative corrections must be invariant under duality, $\delta \mathcal{L}^{(4)}=0$, in order for the theory to respect duality symmetry.

\subsection{More on Four-Derivative $\mathcal{N}=2$ Invariants with Constant Scalars}\label{sec:appendix:invariants}
We can provide a few more calculational details to the discussion in section \ref{sec:duality:fourder}, where the allowed four-derivative  duality-invariant terms are found for $\mathcal{N}=2$ supergravity with constant scalars, with the result (\ref{eq:duality:fourderinvfinal}) that there are only five such independent terms on-shell.

In section \ref{sec:duality:fourder}, it was explained that the only duality-invariants containing at most two symplectic vectors and at most four derivatives are given by (leaving out the zero-derivative symplectic invariant $\bX\cdot \bbX$, which is simply constant)
\begin{align}
\label{eq:appduality:1der} &T^+_{\mu\nu} = -\frac{i\kappa^2}{2\pi} \bF^+_{\mu\nu}\cdot \bbX~, & && && \\
\label{eq:appduality:2der} &\calI_{2\mu\nu\rho\sigma}= \bF^+_{\mu\nu}\cdot \bF^-_{\rho\sigma}~, && \nabla_\rho T_{\mu\nu}^+ = -\frac{i\kappa^2}{2\pi}\nabla_\rho \bF^+_{\mu\nu}\cdot \bbX, && R_{\mu\nu\rho\sigma}~, &\\
 \label{eq:appduality:3der} &\nabla_\lambda \bF^+_{\mu\nu}\cdot \bF^+_{\rho\sigma}~,&& \nabla_\lambda \bF^+_{\mu\nu}\cdot \bF^-_{\rho\sigma}~,&& \nabla_\rho\nabla_\sigma \bF^+_{\mu\nu}\cdot \bbX~, &\\
 \label{eq:appduality:4der}& \nabla_\lambda \bF^+_{\mu\nu}\cdot \nabla_\omega \bF^+_{\rho\sigma}~, && \nabla_\lambda\nabla_\omega\bF^+_{\mu\nu}\cdot \bF^-_{\rho\sigma}~,&& \nabla_\lambda\bF^+_{\mu\nu}\cdot \nabla_\omega \bF^-_{\rho\sigma}~, && \nabla_\lambda\nabla_\omega\bF^+_{\mu\nu}\cdot  \bF^-_{\rho\sigma}~,
\end{align}
and their complex conjugates. Note that each line gives the invariants at a given order in derivatives (from one to four derivatives).

To now find allowed four-derivative terms on-shell, we should multiply the above terms together in such a way that we get a four-derivative term, and then contract Lorentz indices to form a Lorentz scalar. We will use the following principles to determine such terms that are allowed and independent:
\begin{itemize}
 \item \emph{Respecting $U(1)_R$ symmetry:} Under the $\mathcal{N}=2$ global $U(1)_R$ symmetry, $\bX$ and $\bbX$ carry opposite charges ($\bF^{\pm}_{\mu\nu}$ is uncharged). Since the four-derivative Lagrangian should respect the $U(1)_R$ symmetry, any allowed term should have vanishing total $U(1)_R$ charge.
 \item \emph{Discarding total derivatives:} If two allowed terms $T_1, T_2$ are related by a total derivative, $T_1=T_2+\nabla_\mu T_3^\mu$, then we consider $T_2$ equivalent to $T_1$ (and we can discard one of these two terms). This is because we are interested in the independent terms that can appear in a Lagrangian, so we are allowed to add total derivatives to these terms at will.
 \item \emph{Using two-derivative equations of motion:} When the scalars are constant, the two-derivative Einstein equations (\ref{eq:twoderEinstein}) can be seen to equate $R_{\mu\nu}$ to $\calI_{2\lambda(\mu\nu)}^{\ \ \ \ \ \ \lambda}$ of (\ref{eq:appduality:2der}), so we will freely interchange the two and use the relation to eliminate any terms containing $\calI_{2\lambda(\mu\nu)}^{\ \ \ \ \ \ \lambda}$ in favor of $R_{\mu\nu}$. A consequence is also that $R=0$. We will also use the two-derivative Bianchi identity and equations of motion (\ref{eq:twoderFEOM}), (\ref{eq:twoderFBI}) for the vectors, which allows us to set $\nabla_\mu \bF^{+\mu\nu} = \nabla_\mu\bF^{-\mu\nu}$.
 \item \emph{(Anti-)self-duality of $\bF^{\pm}_{\mu\nu}$:} Finally, when contracting Lorentz indices to form a Lorentz scalar, we will use the (anti-)self-duality properties of $\bF^{\pm}_{\mu\nu}$ intensively to relate different ways of contracting Lorentz indices to each other. This will drastically reduce the number of independent four-derivative Lorentz scalars we can construct, as many different contractions of Lorentz indices can often be shown to be equal using these (anti-)self-duality properties. We will also allow ourselves to keep in mind the explicit form of $G^+_{I\,\mu\nu}$ given in (\ref{eq:duality:Gonshell}) in terms of $F^{+I}_{\mu\nu}$.
 \end{itemize}
 
 We now proceed systematically to investigate all possible four-derivative Lorentz scalar terms that we can write down using the above principles:
 \begin{itemize}
  \item We can take a single term from the four quantities in line (\ref{eq:appduality:4der}) and contract Lorentz indices to obtain a four-derivative Lorentz scalar. First of all, it is obvious that we can ignore the second and fourth terms in (\ref{eq:appduality:4der}) as they are equivalent to the third and first term, respectively. Using self-duality of $\bF^+_{\mu\nu}$ and the explicit form (\ref{eq:duality:Gonshell}) of $G^+_{I\, \mu\nu}$ in the vector $\bF^+_{\mu\nu}$, it can be shown that there are actually no non-zero contractions of the first term in (\ref{eq:appduality:4der}). Finally, there is only one independent non-zero contraction of the third term, given by: $\nabla_\mu \bF^{+\mu\nu}\nabla_\rho \bF^{-\rho}_{\ \ \mu}$, but we can use the Bianchi identity and equations of motion to relate $\nabla_\rho \bF^{-\rho}_{\ \ \mu}=\nabla_\rho \bF^{+\rho}_{\ \ \mu}$, so that this term will also vanish.
  \item We can take a quantity from line (\ref{eq:appduality:3der}) and multiply it by (\ref{eq:appduality:1der}). However, we can use total derivatives to relate any such resulting term to a term that is a product of two quantities from (\ref{eq:appduality:2der}), so there are no such independent terms.
  \item We can take two quantities from line (\ref{eq:appduality:2der}) and multiply them together, contracting Lorentz indices. The second quantity in (\ref{eq:appduality:2der}) is charged under $U(1)_R$ and can be multiplied by its complex conjugate to obtain a $U(1)_R$ invariant term. There is one independent way of forming a Lorentz scalar in this way:
  \be \nabla_\mu T^{+\mu\nu} \nabla_\rho T^{-\rho}_{\ \ \ \nu}~.\ee
  We can also multiply the first or third quantities from (\ref{eq:appduality:2der}) amongst themselves. Using the Einstein equations of motion and (anti-)self-duality properties, we can see that there are only two independent such terms:
  \be R_{\mu\nu}R^{\mu\nu}~, \quad R_{\mu\nu\rho\sigma} R^{\mu\nu\rho\sigma}~.\ee
  \item We can take a quantity from line (\ref{eq:appduality:2der}) and multiply it twice with (\ref{eq:appduality:1der}). We must take care that the resulting term is $U(1)_R$ invariant. Then, again using the Einstein equations of motion and (anti-)self-duality properties, we can conclude there is only one independent such term:
  \be R_{\mu\nu} T^{+\mu}_{\ \ \  \rho} T^{-\nu\rho}~. \ee
  \item Finally, we can multiply (\ref{eq:appduality:1der}) or its complex conjugate with itself four times. We must take a $U(1)_R$ invariant term, of course, and (anti-)self-duality properties tell us there is only one such term:
  \be T^{-\rho}_{\ \mu} T^{-\mu\nu}T^{+\ \sigma}_{\ \nu}T^{+}_{\rho\sigma}~.\ee
 \end{itemize}
 Putting everything together, we see we have obtained a total of five independent terms, as given in (\ref{eq:duality:fourderinvfinal}).

\bibliographystyle{jhep}
\bibliography{nonrenormthm}
\end{document}